\documentclass[a4paper,11pt]{article}
\pdfoutput=1 
\usepackage{jheppub} 

  \usepackage{etoolbox}
  \patchcmd{\maketitle}{\@fpheader}{}{}{}

\usepackage[T1]{fontenc} 
\usepackage[utf8]{inputenc} 
\usepackage{microtype} 
\usepackage{mdframed} 
\usepackage{graphicx}
\usepackage{todonotes} 

\usepackage{amsmath}
\usepackage{amsfonts}
\usepackage{amssymb}
\usepackage{stmaryrd}
\usepackage{mathtools}
\usepackage{mathrsfs}
\usepackage{scalerel}
\usepackage{mleftright} \mleftright
\usepackage{tensor} 
\usepackage{braket} 
\usepackage{mathrsfs} 
\usepackage{bbold}
\usepackage{color}
\usepackage{braket}
\usepackage{slashed}

\newcommand{\comment}[1]{}

\providecommand*{\pd}{\mathop{}\!\partial}
\renewcommand*{\pd}{\mathop{}\!\partial}

\providecommand*{\cd}{\mathop{}\!\nabla}
\renewcommand*{\cd}{\mathop{}\!\nabla}

\renewcommand{\vec}[1]{\mathbf{#1}}

\providecommand{\Lc}{{\mathcal{L}}}
\renewcommand{\Lc}{{\mathcal{L}}}

\providecommand{\H}{{\mathcal{H}}}
\renewcommand{\H}{{\mathcal{H}}}

\DeclareMathAlphabet{\mathfs}{U}{rsfs}{m}{n}                     %

\newcommand{\h}{\hat} 
\newcommand{\D}{\dagger}

\newcommand{\n}{\nonumber}
\newcommand{\be}{\nopagebreak[3]\begin{equation}}
\newcommand{\ee}{\end{equation}}
\newcommand{\bee}{\nopagebreak[3]\begin{equation*}}
\newcommand{\eee}{\end{equation*}}
\newcommand{\ba}{\nopagebreak[3]\begin{eqnarray}}
\newcommand{\ea}{\end{eqnarray}}
\newcommand{\baa}{\nopagebreak[3]\begin{eqnarray*}}
\newcommand{\eaa}{\end{eqnarray*}}

\title{UNRUH EFFECT\\
\vspace{0.5cm}
Introductory Notes to\\ Quantum Effects for Accelerated Observers}

\author[1,2]{Ernesto Frodden}
\author[3]{and Nicol\'as Vald\'es}

\affiliation[1]{Facultad de F\'{i}sica, Pontificia Universidad Cat\'olica de Chile, Casilla 306, Santiago, Chile}
\affiliation[2]{Centro de Estudios Cient\'ificos (CECs), Av. Arturo Prat 514, Valdivia, Chile}
\affiliation[3]{Departamento de F\'isica, FCFM, Universidad de Chile, Blanco Encalada 2008, Santiago, Chile}

\emailAdd{efrodden@gmail.com}
\emailAdd{n.valdes.meller@gmail.com}

\abstract{
These pedagogical notes are dedicated to a derivation of the Unruh effect. There is special emphasis on the transparency of the arguments and the exhibition of detailed calculations. We assume the reader has a  basic knowledge of quantum mechanics and  relativity.  The notes can hopefully be used in a self-contained way by advanced undergraduate or beginning graduate students. 
}

\begin{document}

\maketitle
\flushbottom


\begin{flushright}
{\it To heat up a cup of tea you may accelerate it...}
\end{flushright}

\section{Introduction}
\label{sec:introduction}

An accelerated observer measures her surrounding fields in excited thermal states. A simple thermometer carried  by the observer marks a temperature directly proportional to the acceleration she is moving with. The more she accelerates, the higher the temperature.

This is the surprising conclusion W. Unruh reached  in 1976 \cite{Unruh:1976db}, just a couple of years after S. Hawking concluded that black holes are not eternal in the sky, but rather  they evaporate and shrink. 
As strange as it may seem at the beginning, both phenomena are intimately related. Both exploit the existence of causal horizons on spacetime and use the same mathematical tools to describe quantum fields there. In a wider view both of them rest on physics pillars still solid today: a classical description for spacetime and a quantum description for matter given by the theories of general relativity and quantum field theory, respectively. 

The Unruh effect is a first step to introduce one-self to the study of quantum field theory in curved spacetime and it turns out that this theory is  among the firmest current ways to deal with the problem of quantum gravity. Thus, 42 years later the Unruh effect is still a cornerstone in theoretical physics. It is an example of physics at its best. Standing up over well stablished theories, the Unruh effect is a prediction of a fantastic phenomenon that at the same time can be used to question the limits of the theories themselves.

These notes are meant to be pedagogical; most computations are performed in detail, and many preliminaries are discussed thoroughly. We hope that with a basic knowledge of quantum mechanics and relativity the concepts outlined will be comprehensible. 

Ideally, the notes will help advanced undergraduate or beginning graduate students delve into this interesting effect; to this end we have included some simple exercises at the end of the notes. Our overall purpose is to help  bring community attention and to soften the introduction for newcomers into what we think is still a window to the problem of quantum gravity.

We do not discuss more technical problems of quantum field theory in curved spacetime, since the Unruh effect is studied in flat spacetime. We do, however, discuss and outline the general idea of quantum field theory in curved spacetime for completeness. 

The notes are organized as follows. In the preliminaries, after a very brief review of what the canonical quantization procedure entails, we discuss the quantization of the harmonic oscillator, but introducing a point of view not typically considered in introductory quantum mechanics courses. For this reason it is recommended that even readers familiar with the quantum harmonic oscillator read this section. After the oscillator, we give a detailed review of the quantization of a free scalar field in flat spacetime, and then in curved spacetime.  

Following the preliminaries, we delve into the Unruh effect. We begin with a brief review of spacetimes for accelerated observers: Rindler spacetime. We then find the positive frequency modes of a Rindler observer, and see how to construct similar modes, but covering all of spacetime. Using these new modes, we compare the vacuum of Rindler observers with that of inertial observers, arriving at Unruh's famous result.  We explain in detail a core elegant argument given by Unruh, discussed in section \ref{elegant}, something we could not find elsewhere. The argument, though clever, is subtle and not completely transparent at a first glance; we hope our attempt here has made it more accessible.

The notes are based primarily on S. Carroll's book on general relativity \cite{Carroll:2004st} and T. Jacobson's review on quantum field theory in curved spacetime (QFT-CS) \cite{Jacobson:2003vx}. We also considered the review by Crispino et al \cite{Crispino:2007eb}, Birrell and Davies' book \cite{Birrell:1982ix}, and Unruh's original 1976 paper \cite{Unruh:1976db}.

\section{Preliminaries}
\label{preliminaries}

In this section we outline the procedure for canonical quantization and apply it to three different cases:  The harmonic oscillator, a scalar field in flat spacetime, and a scalar field in curved spacetime. 

What is canonical quantization? It is simply a prescription for quantizing classical systems based on their canonically conjugate variables. Usually one starts considering a Lagrangian description of a classical system (in the example below the harmonic oscillator): $L(q^i,\dot{q}^i,t)$ where $\dot{q}=\frac{dq}{dt}$. The equations of motion are obtained through the variational principle. Varying the action $S=\int dt L$ with respect to each $q^i$, choosing boundary conditions that guarantee the vanishing of the boundary term, and demanding this variation be zero yields the Euler-Lagrange equation
\be
\frac{\pd L}{\pd q^i} - \frac{d}{dt} \frac{\pd L}{\pd \dot{q}^i} = 0. 
\ee
We define the conjugate momentum of $q^i$ by $p_i\equiv\pd L/\pd \dot{q}^i$. The Hamiltonian is then defined $H\equiv p_i\dot{q}^i-L$, where $\dot q^i=\dot q^i(q_j,p_j)$. Repeated indices are summed over and note that $q_i=q^i$. In the Hamiltonian formalism $q_i$ and $p_i$ are called {\it canonical variables} as they satisfy the Poisson bracket relations 
\be
\{q_i,p_j\}=\delta_{ij}, \quad \{q_i,q_j\}=\{p_i,p_j\}=0,
\ee
where the Poisson bracket is 
\bee
\{f,g\} \equiv \frac{\pd f}{\pd q^i} \frac{\pd g}{\pd p_i} - \frac{\pd f}{\pd p_i}\frac{\pd g}{\pd q^i}. 
\eee
The Poisson bracket is defined in a more formal way as a bilinear anticommutative binary operation on functions of phase space and time, that satisfies Leibniz's rule and the Jacobi identity. 

In the quantization procedure the classical variables are promoted to Hermitian operators acting on a Hilbert space. The commutator of two operators representing canonically conjugate variables satisfies the same properties as the Poisson bracket. 
The physical system is no longer described by canonical variables, but instead by states defined on the Hilbert space. Therefore, the Poisson bracket relations between $q_i$ and $p_j$ are promoted to commutation relations between Hermitian operators $\h{q}_i$ and $\h{p}_j$ acting on the Hilbert space such that
\be
\hspace{1cm} [\h{q}_i,\h{p}_j] = i\hbar \delta_{ij},\quad [\h{q}_i,\h{q}_j] = [\h{p}_i,\h{p}_j] = 0, 
\ee
where we introduce $i\hbar$ in the commutator.
 
To be concrete let us carry out this procedure for the harmonic oscillator.

\subsection{Quantization of the Harmonic Oscillator}

 The reason we consider the harmonic oscillator is that this system is the one-dimensional degree of freedom counterpart of a free scalar field; many of the concepts defined for the oscillator carry over directly to scalar fields and from them to more general field theories. The Lagrangian for our system is
\be
L = \frac{1}{2}m\dot{x}^2 -\frac{1}{2}m\omega^2 x^2. \nonumber
\ee
The conjugate momentum is $p=m\dot{x}$. 
Applying the Euler-Lagrange equation to the above Lagrangian, we find that $x$ satisfies 
\be 
\ddot{x}+\omega^2 x=0. \label{motion1} 
\ee  
Now we apply the canonical quantization procedure. In contrast with the usual {\it Schr\"odinger picture} where states evolve in time here we use the {\it Heisenberg picture} where states are fixed and evolution is encoded in the operators. We turn both position and momentum into operators, so the equation of motion (\ref{motion1}) becomes an equation for operators.  Additionally, we impose the canonical commutation relations
\be
[\h{x}(t),\dot{\h{x}}(t)] = \frac{i\hbar}{m}. \label{commutation} 
\ee
The general operator solution to (\ref{motion1}) has the form 
\be
\h{x}=\h{a}f(t)+\h{a}^{\dagger}f^*(t) \label{x},
\ee
where $\h{a}$ is some (constant, non-Hermitian) operator, and $f$ satisfies (\ref{motion1}). Note that the coefficient of the second term is the adjoint of the first coefficient (and not some arbitrary operator); this is to ensure that $\h{x}$ is Hermitian. 

The Hamiltonian operator of our system becomes 
\be 
\h{H} = \frac{1}{2}m\dot{\h{x}}^2 +\frac{1}{2}m\omega^2\h{x}^2.  \label{H}
\ee

Now, we introduce some formalism which will be useful later. We define an inner product on the space of solutions to (\ref{motion1}) in the following way 
\be 
\langle f,g \rangle = \frac{im}{\hbar} \left(f^*\pd_t g- (\pd_t f^*)g\right). \label{inner1} 
\ee
This is a reasonable inner product, since it is conserved when $f$ and $g$ solve (\ref{motion1}) (i.e. its time derivative is zero, as you're asked to show in exercise \ref{timederivative}). Using this notation, the commutation relation (\ref{commutation}), and plugging in the general solution (\ref{x}), takes an interesting form 
\begin{align*}
[\h{x}(t),\dot{\h{x}}(t)] &= [\h{a}f+\h{a}^{\dagger}f^*, \h{a}\pd_t f+\h{a}^{\dagger}\pd_t f^*] \\
&= [\h{a},\h{a}^{\D}] f\pd_t f^* + [\h{a}^{\D},\h{a}] f^* \pd_t f\\
&= [\h{a},\h{a}^{\D}] (f\pd_t f^* - f^*\pd_t f) = \frac{i\hbar}{m}. 
\end{align*}
Thus using the new notation, we have 
\be
\langle f,f\rangle [\h{a},\h{a}^{\dagger}]=1. \label{positivefreq} 
\ee
Here we will assume that we can choose $f$ such that $\langle f,f\rangle>0$. For a reason that will be clear in a moment this $f$ will be called a positive frequency mode. This doesn't seem very important now, but identifying positive frequency modes will be fundamental when we quantize fields in curved spacetimes. After ensuring that $\langle f,f\rangle>0$, we can also normalize $f$ such that $\langle f,f\rangle=1$. So we're left with 
\be
[\h{a},\h{a}^{\dagger}]=1, \label{great} 
\ee
which will turn out to be an extremely useful relation. Additionally,
with these definitions we have
\be
\h{a}=\braket{f,\h{x}}\label{hata},
\ee
explicitly
\begin{align*}
\braket{f,\h{x}} &= \frac{im}{\hbar}\left(f^* \dot{\h{x}}-\dot{f}^*\h{x}\right)\\
&=\frac{im}{\hbar}\left(\h{a}(f^*\dot{f}-\dot{f}^*f)+\h{a}^{\D}(f^*\dot{f}^*-f^*\dot{f}^*)\right) \\
&=\h{a}\braket{f,f}\\
&=\h{a}.
\end{align*}
To go from the first to the second line, we used the (\ref{x}) expansion for $\h{x}$. For the next step we used the definition of the inner product, and for the last step we used that $f$ is a normalized positive frequency mode. 

Let's make the discussion more concrete by noting that  $f(t)=Ae^{-i\omega t}$ solves (\ref{motion1}) for scalars (verify this). The factor $A$ that accompanies $e^{-i\omega t}$ will be equal to $\sqrt{\hbar/2m\omega}$ for $f$ to be normalized. Let's verify this statement  
\ba
\langle f, f \rangle &=& \frac{im}{\hbar} \left(f^*\pd_t f- (\pd_t f^*)f\right)\n\\
&=&\frac{im}{\hbar} \left(\sqrt{\frac{\hbar}{2m\omega}}e^{i\omega t}\cdot (-i\omega) \sqrt{\frac{\hbar}{2m\omega}}e^{-i\omega t} - (i\omega)\sqrt{\frac{\hbar}{2m\omega}}e^{i\omega t}\cdot\sqrt{\frac{\hbar}{2m\omega}}e^{-i\omega t}  \right)\n\\
&=&1,\n
\ea
so indeed, $f$ is a normalized positive frequency mode.  In particular, it satisfies $i\partial_t f = +\omega f$ with $\omega>0$, which is another way to characterize a positive frequency mode.

Now let's get back to the commutation relation (\ref{great}) (which we arrived at by assuming we could have an $f$ such that $\langle f,f\rangle=1$). This relation is extremely important. To see why, we can expand the Hamiltonian (\ref{H}) using (\ref{x}) 
\begin{align*}
\h{H}&=\frac{1}{2}m(\h{a}\dot{f}+\h{a}^{\D}\dot{f}^*)^2+\frac{1}{2}m\omega^2(\h{a}f+\h{a}^{\dagger}f^*)^2\\
&=\frac{1}{2}m\left\{(\dot{f}^2+\omega^2f^2)\h{a}\h{a}+(\dot{f}^2+\omega^2 f^2)^* \h{a}^{\dagger}\h{a}^{\dagger} +(|\dot{f}|^2+\omega^2|f|^2)(\h{a}\h{a}^{\dagger}+\h{a}^{\dagger}\h{a})\right\}.\nonumber
\end{align*}

Now recall that $f$ satisfies the equation of motion, so the first two terms in the above expression are zero. 

Plugging in the explicit form $f=Ae^{-i\omega t}$ for the last term, we have 
\be
\h{H}=\frac{1}{2}\hbar \omega(\h{a}\h{a}^{\dagger}+\h{a}^{\dagger}\h{a}), \nonumber
\ee 
since $|f|^2=A^2=\hbar/2m\omega$.  Note that $\h{a}\h{a}^{\D}+\h{a}^{\D}\h{a}=[\h{a},\h{a}^{\D}]+2\h{a}^{\D}\h{a}$, so we have the following form for our Hamiltonian 
\be
\h H=\hbar\omega\left(\h{a}^{\dagger}\h{a}+\frac{1}{2}\right). 
\ee

Given this  Hamiltonian, we can deduce that $[\h{H},\h{a}]=-\hbar\omega\h{a}$, and that $[\h{H},\h{a}^{\D}]=\hbar\omega\h{a}^{\D}$. And here we begin to see what's great about the commutation relation (\ref{great}). We have that for some eigenstate $\ket{n}$ of $\h H$, with energy $E_n$, 
\bee
\h{H} (\h{a}\ket{n}) = (\h{a}\h{H}+[\h{H},\h{a}])\ket{n} = (E_n\h{a}-\hbar\omega\h{a})\ket{n}= (E_n-\hbar\omega)\h{a}\ket{n}. 
\eee
This means that if $\ket{n}$ is an eigenstate of $\h{H}$ with energy $E_n$, then $\h{a}\ket{n}$ is an eigenstate of $\h{H}$ with energy $E_n-\hbar\omega$. Similarly, it can be shown that $\h{a}^{\D}\ket{n}$ is an eigenstate of $\h{H}$ with energy $E_n+\hbar\omega$.

So we can deduce that $\h{a}\ket{n}=\sqrt{n} \ket{n-1}$, and $\h{a}^{\dagger}\ket{n}=\sqrt{n+1}\ket{n+1}$ (with the factors appearing due to normalization). For this reason we call $\h{a}^{\dagger}$ and $\h{a}$ the raising and lowering operators (they raise/lower the energy of the system). You should verify that the operator $\h{n}=\h{a}^{\D}\h{a}$, called the number operator, satisfies $\h{n}\ket{n}=n\ket{n}$. That is, it ``counts'' the energy levels of the oscillator. Now, recall that the harmonic oscillator is subject to a positive quadratic potential, so for normalizable solutions, we have positive energy. This means that we can't lower states forever; there must be some state $\ket{0}$ such that $\h{a}\ket{0}=0$. This is the smallest-energy state, or the {\it ground state} of the Hamiltonian. It's easy to see that $\h{H}\ket{0}=\frac{1}{2}\hbar\omega\ket{0}$: the ground state of the oscillator has an energy different from zero called the ``zero-point energy.'' From this state $\ket{0}$, we can construct a basis for the Hilbert space of our system using the raising operator: $\ket{n}=\frac{1}{\sqrt{n!}}(\h{a}^{\dagger})^n\ket{0}$.  In the following section we apply analogous versions of the concepts we've developed here, in order to describe quantum \textit{fields}.

\subsection{Quantization of a Scalar Field}

Let's first introduce some conventions. We use the metric signature $(-1,+1,+1,...)$ in $n$ dimensions, and set $c=G=\hbar=1$. Repeated indices are summed over, e.g. $x_{\mu}p^{\mu}=\sum_{\mu=0}^{n-1} x_{\mu}p^{\mu}$. 
Vectors are denoted by their components. For example, $k^{\mu}$ corresponds to the vector $k=k^{\mu}\partial_\mu$ (with $\mu=0,\dots, n-1$), and we denote the spatial part of vectors in boldface, ${\vec k}=k^{i}\partial_i$ (with $i=1,\dots, n-1$). The tensor $\eta_{\mu\nu}=\text{diag}(-1,+1,+1,...)$ corresponds to the Minkowski metric, and $g_{\mu\nu}$ to a generic metric.

Canonical quantization of a scalar field theory is a similar story to what we did for particles, but we have a field $\phi(x^{\mu})$ and its associated momentum $\pi(x^{\mu})$ instead of position and momentum to begin with. We'll consider a free real scalar field because it is enough for illustrating the theory without unnecessary clutter; extending what we'll do to physically realistic fields (like spin $\frac{1}{2}$ fields or the electromagnetic field) isn't too complicated.
What does this field represent? This field doesn't have a direct physical meaning, and that can be hard to grasp at first. We don't really worry about that for now, we'll start with an action for the scalar field, find equations of motion for it, quantize the system, and from there interpret results that are generic for other fields too. 

To quantize a field theory, we have practically the same procedure as for particles. 
First we recall what is done in classical field theory: Given a Lagrangian density $\Lc(\phi,\pd_{\mu}\phi)$, the equation of motion for $\phi$ is given by varying the action $S=\int d^4x \Lc$ with respect to $\phi$, demanding that this variation be zero, and properly getting rid of the boundary terms. This yields the Euler-Lagrange equation for fields
\be
\frac{\pd \Lc}{\pd \phi} - \pd_{\mu}\frac{\pd \Lc}{\pd (\pd_{\mu}\phi)}=0.
\ee
We have the field conjugate momentum to $\phi(x^{\mu})$ defined by $\pi(x^{\mu})\equiv\pd \Lc/\pd \dot{\phi}$ with $\dot \phi\equiv \partial_t\phi$. 
The Hamiltonian density is $\H\equiv\pi \dot{\phi}-\Lc$, and the Hamiltonian $H=\int d^{n-1}x \H$. 

To quantize, we promote the scalar fields $\phi$ and $\pi$ to field operators ($\phi(x^\mu)$ and $\pi(x^\mu)$ are operators valued at each spacetime point $x^\mu$). We won't put hats on the fields, to simplify notation. But remember, they become operators.  

Then we impose canonical commutation relations 
\be \label{commutation2}
[\phi(t,\vec{x}),\pi(t,\vec{y})]=i\delta^{(n-1)}(\vec{x}-\vec{y}), \quad [\phi(t,\vec{x}),\phi(t,\vec{y})]=[\pi(t,\vec{x}),\pi(t,\vec{y})]=0.
\ee
Note that these commutation relations are taken at \textit{equal times}. 

Having outlined the starting procedure for canonical quantization of fields, we proceed with a specific example. Consider a field $\phi$ of mass $m$ characterized by the following Lagrangian density 
\be
\Lc=-\frac{1}{2}\eta^{\mu\nu}\pd_{\mu}\phi\pd_{\nu} \phi -\frac{1}{2}m^2\phi^2. \label{lagrangian}
\ee 
Note that the Lagrangian density is a Lorentz-scalar. Just like with the oscillator, to find the equation of motion we can vary the action (and require that its variation be an extremum), or just use the Euler Lagrange equation. We have to be a little careful in doing this,
\bee
\frac{\pd \Lc}{\pd \phi}= -m^2\phi, \hspace{1cm} \frac{\pd \Lc}{\pd (\pd_{\alpha}\phi)}= \frac{\pd}{\pd (\pd_{\alpha}\phi)} \left(-\frac{1}{2}\eta^{\mu\nu}\pd_{\mu}\phi\pd_{\nu}\phi\right)= -\eta^{\mu\alpha} \pd_{\mu}\phi,
\eee
since $\pd (\pd_{\mu}\phi)/\pd (\pd_{\alpha}\phi)= \delta^{\alpha}_{\mu}$. 
So we arrive at the Klein-Gordon equation
\be
(\square -m^2) \phi=0 \label{KG},
\ee
where $\square \equiv \eta^{\mu\nu}\pd_{\mu}\pd_{\nu}$. Then we compute the conjugate momentum to $\phi$ 
\bee
\pi(x^{\mu})=\frac{\pd \Lc}{\pd \dot{\phi}} = \dot{\phi}, 
\eee
and the Hamiltonian density 
\be
\H = \pi\dot{\phi}-\Lc= \frac{1}{2} \pi^2+\frac{1}{2} (\cd\phi)^2+\frac{1}{2}m^2\phi^2.
\ee
Finally, to quantize the theory we promote $\phi$ and $\pi$ to operators and impose the canonical commutation relations (\ref{commutation2}).

Now getting back to (\ref{KG}), we do something similar to what we did with the oscillator. There, we expanded $\h{x}$ as $\h{a}f+\h{a}^{\D}f^*$, with $f$ a solution to the $\h{x}$ equation of motion, and $\h{a}$ a constant operator. Here since we have a field we use the following expansion 
\be
\phi(t,\vec{x}) = \int d^{n-1}k\left( \h{a}_{\vec{k}}\varphi_{\vec{k}}(t,\vec{x})+\h{a}^{\D}_{\vec{k}}\varphi_{\vec{k}}^*(t,\vec{x})\right), \label{expansion1}
\ee
where the $\varphi_{\vec{k}}$ are a complete set of solutions to the scalar version of (\ref{KG}), parametrized by some vector $\vec{k}$, and $\h{a}_{\vec{k}}$ are constant operators attached to each solution. 
This expression is completely analogous to what we had in (\ref{x}), except that we now have a continuous set of functions spanning our solution space, rather than just one function and its conjugate. And, of course, we're solving (\ref{KG}) now instead of (\ref{motion1}). 

This is just an expansion in a basis, which isn't special---it can always be done. 
We assume that this is an orthonormal basis where the $\varphi_{\vec{k}}$ are chosen as {\it positive frequency modes} (the $\varphi_{\vec{k}}^*$ are negative frequency modes). To give a precise meaning to this last statement we must first define an inner product on the solution space to (\ref{KG}) 
\be
\langle f,g\rangle \equiv i \int_{\Sigma} d^{n-1}x (f^* \partial_t g - g\partial_t f^*). \label{inner2} 
\ee
The integral is over a constant time hypersurface $\Sigma$, which are particular Cauchy surfaces for (\ref{KG}).
This is analogous to the scalar product (\ref{inner1}) we defined on the solution space to the harmonic oscillator equation. It is also a reasonable inner product, since it is independent of the Cauchy surface it is taken over.  Let us verify this using covariant notation. 

Consider a volume $M$ contained in a surface $\Sigma$ that can be decomposed as $\Sigma=\Sigma_1 \cup \Sigma_2$, then, Stokes's theorem states that  
\be
\int_{\Sigma_1}d\Sigma^\mu j_\mu+\int_{\Sigma_2}d\Sigma^\mu j_\mu=\int_{M}dV \nabla^\mu j_\mu, \label{stokes}
\ee
with $d\Sigma^{\mu}$ the covariant surface element on $\Sigma_1$ and $\Sigma_2$. In flat spacetime with Cartesian coordinates the volume element on $M$ reduces to $dV=d^nx$, and the covariant derivative is a normal partial derivative $\nabla_\mu=\partial_\mu$. The vector $j_\mu$ is arbitrary but if we consider $j_\mu=i(f^*\partial_\mu g-g\partial_\mu f^*)$ we have that $\nabla^\mu j_\mu=i(f^*\square g-g\square f^*)=0$, so the r.h.s. vanishes due to the equations of motion.
Also in Cartesian coordinates $d\Sigma^{\mu}=d^{n-1}x\ n^{\mu}$, where $n^{\mu}$ are the components of a unit vector pointing outside the volume $M$. If $\Sigma_2$ is in the future of $\Sigma_1$, the normal vector is $\left.n^\mu\right|_{\Sigma_2}=(1,0,\cdots)$ and $\left.n^\mu\right|_{\Sigma_1}=(-1,0,\cdots)$.  Therefore, on $\Sigma_2$ we have $d\Sigma^\mu j_\mu=d^{n-1}x j_0=id^{n-1}x (f^* \partial_t g - g\partial_t f^*)$ (with the opposite sign for $\Sigma_1$). Then the vanishing of (\ref{stokes}) implies\footnote{\label{foot} There is a technical point worth noting: Two constant time surfaces do not completely enclose a volume, instead $M=\Sigma_1\cup \Sigma_2\cup\tau$ with $\tau$ a third surface that can be thought of as the surface of a higher dimensional cylinder of infinite radius (whose top and bottom bases are $\Sigma_1$ and $\Sigma_2$). Instead of considering this surface explicitly we should restrict the space of solutions such that the considered solutions vanish on $\tau$. This interplay between boundary and space of solutions is a recurring difficulty to define an inner product on curved spacetimes.} 
\be
\langle f,g\rangle_{\Sigma_1}=i \int_{\Sigma_1} d^{n-1}x (f^* \partial_t g - g\partial_t f^*)=i \int_{\Sigma_2} d^{n-1}x (f^* \partial_t g - g\partial_t f^*)=\langle f,g\rangle_{\Sigma_2}\n
\ee
and thus $\langle f,g\rangle$ does not depend on the constant time (Cauchy) surface we choose.

 
Having justified our choice of (\ref{inner2}), we now clarify that
when we say $\varphi_{\vec{k}}$ is a normalized positive frequency mode, we mean that $\braket{\varphi_{\vec{k}},\varphi_{\vec{q}}}=\delta^{(n-1)}(\vec{k-q})$. Analogously, for negative frequency modes we have $\braket{\varphi_{\vec{k}}^*,\varphi_{\vec{q}}^*}=-\delta^{(n-1)}(\vec{k-q})$, and the modes are orthogonal, i.e. $\braket{\varphi_{\vec{k}},\varphi_{\vec{q}}^*}=\braket{\varphi_{\vec{k}}^*,\varphi_{\vec{q}}}=0$. This definition is analogous to $\braket{f,f}=1$ for the harmonic oscillator.

Now recall the ``great'' commutation relation for the oscillator, with the raising and lowering operators. We didn't use the names $\h{a}_{\vec{k}}$ and $\h{a}^{\D}_{\vec{k}}$  for the coefficients in the field expansion (\ref{expansion1}) accidentally. It'll turn out that we have similar commutation relations for these operators, and we can later attach a physical interpretation to them. So we will work to find $[\h{a}_{\vec{k}},\h{a}_{\vec{q}}^{\D}]$. When we considered the oscillator, the relation $\braket{f,f}=1$ allowed us to arrive at $[\h{a},\h{a}^{\D}]=1$. Now our definition of positive frequency modes, $\braket{\varphi_{\vec{k}},\varphi_{\vec{q}}}=\delta^{(n-1)}(\vec{k-q})$, will give us analogous commutation relations. 



Given that the $\varphi_{\vec{k}}$ are orthonormal, the expansion (\ref{expansion1}) tells us that 
\be
\h{a}_{\vec{k}}=\braket{\varphi_{\vec{k}},\phi}, \label{a_k}
\ee
see (\ref{hata}).
The explicit computation is 
\begin{align*}
\braket{\varphi_{\vec{k}},\phi} &= i \int_{\Sigma} d^{n-1} x (\varphi_{\vec{k}}^*\dot{\phi}-\phi \dot{\varphi_{\vec{k}}}^*) \\
&= i\int_{\Sigma} d^{n-1}x \left( \varphi_{\vec{k}}^* \int d^{n-1}q (\h{a}_{\vec{q}}\dot{\varphi_{\vec{q}}} +\h{a}_{\vec{q}}^{\D} \dot{\varphi_{\vec{q}}}^*) -\dot{\varphi_{\vec{k}}}^* \int d^{n-1}q (\h{a}_{\vec{q}}\varphi_{\vec{q}} +\h{a}_{\vec{q}}^{\D}\varphi_{\vec{q}}^*) \right)\\
&= i \int d^{n-1}q \left(\int_{\Sigma}d^{n-1}x\left\{\h{a}_{\vec{q}}(\varphi_{\vec{k}}^* \dot{\varphi_{\vec{q}}}-\dot{\varphi_{\vec{k}}}^*\varphi_{\vec{q}}) +\h{a}_{\vec{q}}^{\D}(\varphi_{\vec{k}}^*\dot{\varphi_{\vec{q}}}^*-\dot{\varphi_{\vec{k}}}^*\varphi_{\vec{q}}^* )\right\} \right)\\
&=  \int d^{n-1} q \left( \h{a}_{\vec{q}}\braket{\varphi_{\vec{k}},\varphi_{\vec{q}}}+\h{a}_{\vec{q}}^{\D}\braket{\varphi_{\vec{k}},\varphi_{\vec{q}}^*} \right)\\
& = \h{a}_{\vec{k}},
\end{align*}
where we used that the basis is orthonormal. 

 Using (\ref{a_k}), 
\begin{align*}
[\h{a}_{\vec{k}},\h{a}_{\vec{q}}^{\D}] &= [\braket{\varphi_{\vec{k}},\phi},\braket{\varphi_{\vec{q}},\phi}^{\D}] \\
&= \left[\int_{\Sigma} d^{n-1}x \left(\varphi_{\vec{k}}^*(\vec{x})\dot{\phi}(\vec{x})-\dot{\varphi_{\vec{k}}}^*(\vec{x})\phi(\vec{x})  \right),\int_{\Sigma} d^{n-1}y \left(\varphi_{\vec{q}}(\vec{y})\dot{\phi}(\vec{y})-\dot{\varphi_{\vec{q}}}(\vec{y})\phi(\vec{y})  \right)  \right] \\
&=  \int_{\Sigma}d^{n-1}x \int_{\Sigma}d^{n-1}y \left\{-[\pi(\vec{x}),\phi(\vec{y})]\varphi_{\vec{k}}^*(\vec{x})\dot{\varphi_{\vec{q}}}(\vec{y})-[\phi(\vec{x}),\pi(\vec{y})]\dot{\varphi_{\vec{k}}}^*(\vec{x})\varphi_{\vec{q}}(\vec{y}) \right\}\\
& = i \int_{\Sigma}d^{n-1}x \int_{\Sigma}d^{n-1}y \left\{\delta^{(n-1)}(\vec{x-y}) (\varphi_{\vec{k}}^*(\vec{x})\dot{\varphi_{\vec{q}}}(\vec{y})-\dot{\varphi_{\vec{k}}}^*(\vec{x})\varphi_{\vec{q}}(\vec{y}))\right\}\\
&= i\int_{\Sigma} d^{n-1}x \left\{\varphi_{\vec{k}}^*(\vec{x})\dot{\varphi_{\vec{q}}}(\vec{x})-\dot{\varphi_{\vec{k}}}^*(\vec{x})\varphi_{\vec{q}}(\vec{x}))\right\}\\
&= \braket{\varphi_{\vec{k}},\varphi_{\vec{q}}},
\end{align*}
where we've used that $\phi$ and $\pi=\dot{\phi}$ are Hermitian, and applied the canonical commutation relations (\ref{commutation2}). Since our basis is orthonormal, we obtain the simple commutation relation 
\be
[\h{a}_{\vec{k}},\h{a}_{\vec{q}}^{\D}] = \delta^{(n-1)}(\vec{k-q}) \label{commutator}.
\ee
It can similarly be shown that
\be
[\h{a}_{\vec{k}},\h{a}_{\vec{q}}] =[\h{a}_{\vec{k}}^{\D},\h{a}_{\vec{q}}^{\D}] =0.
\ee
These commutation relations will be of great use to us later on. Note that the key of the construction is the definition of an inner product on the solution space. This allows us to deploy the formalism in a rather abstract way, where the equation of motion and its specific solutions may remain generic.

 But now, just like we did with the oscillator, we consider the explicit form of the basis modes to make everything more concrete. Explicit solutions to (\ref{KG}) are 
\be
\varphi_{\vec{k}}(x^{\mu})=A_ke^{i(\vec{k}\cdot\vec{x}-\omega_k t)},  \label{modes1}
\ee
where $\omega_k$ can be positive or negative and satisfies $\omega_k^2=\vec{k}^2+m^2$, and $A_k$ is some constant. 

To convince ourselves that the mode expansion (\ref{expansion1}) is right, let us arrive at it again from a slightly different perspective. We forget momentarily what we did before (\ref{modes1}). 

The modes form a complete basis, so our field $\phi$ can be expanded in these modes
\bee
\phi(x^{\mu}) =  \int d^{n-1} k \left(\h{a}_{\vec{k}} A_ke^{i(\vec{k}\cdot\vec{x}-\omega_k t)}  + \h{b}_{\vec{k}} B_ke^{i(\vec{k}\cdot\vec{x}+\omega_k t)}  \right) .
\eee
We choose $\omega_k$ positive, and divide the solution into two parts corresponding to the positive and negative roots of $\vec{k}^2+m^2$. These parts correspond to the propagating modes in the ${\bf k}$ and $-{\bf k}$ spatial directions, respectively. 
 We attached constant operators to the modes since $\phi$ is an operator.
This expression is a Fourier decomposition of the field $\phi$. 

We can write $\phi$ in a more suggestive way by noting that the integral measure and $\omega_k$ are both invariant under $\vec{k}\to-\vec{k}$, so that the second term can also be written as $\h{b}_{-\vec{k}}B_{-k}e^{-i(\vec{k}\cdot\vec{x}-\omega_kt)}$. If we impose $\phi$ to be Hermitian, then $\h{b}_{-\vec{k}}B_{-{\bf k}}=A_{\bf k}^*\h{a}_{\vec{k}}^{\dagger}$, and finally 
\begin{align}
\phi(x^{\mu}) =  \int d^{n-1} k \left(\h{a}_{\vec{k}} \varphi_{\vec{k}}  + \h{a}_{\vec{k}}^{\D}\varphi^*_{\vec{k}} \right), \label{expansion} 
\end{align}
which is what we had assumed earlier. 

Our choice to keep $\omega_k$ positive is an important one. It gives another meaning to the name positive frequency mode; it is literally a mode with positive frequency. However, keep in mind: The splitting of $\phi$ into positive and negative frequency modes is not special, in principle it's just a mathematical choice related to the coordinates used to describe our differential equation. However, we'll be able to ascribe an appealing physical interpretation to the \textit{coefficients} accompanying  the modes. 

As an aside, you are encouraged to verify that the $\varphi_{\vec{k}}$ we defined are positive frequency modes in the sense of $\braket{\varphi_{\vec{k}},\varphi_{\vec{q}}}>0$, and that their conjugates are negative frequency modes.  

 Now we'll normalize our modes $\varphi_{\vec{k}}$ to determine the constants $A_k$ that accompany them.  We require $\braket{\varphi_{\vec{k}}|\varphi_{\vec{q}}}=\delta^{(n-1)}(\vec{k}-\vec{q})$ for normalization. Plugging (\ref{modes1}) into the inner product definition 
\bee 
\braket{\varphi_{\vec{k}}|\varphi_{\vec{q}}} = i \int d^{n-1}x \left(-i\omega_k A_q^*A_ke^{i(\vec{k}-\vec{q})\cdot\vec{x}}e^{i(\omega_q-\omega_k)t} -i\omega_q A_q^*A_ke^{i(\vec{k}-\vec{q})\cdot\vec{x}}e^{i(\omega_q-\omega_k)t}  \right),
\eee
but the Fourier expansion for the Delta function is 
\bee
\int \frac{e^{i\vec{k}\cdot\vec{x}} }{(2\pi)^{n-1}} d^{n-1}x  = \delta^{(n-1)} (\vec{k}).  
\eee
Comparing this (which we will use repeatedly throughout these notes) with the above expression, we conclude that for normalization $A_k$ must satisfy
\bee
\delta^{(n-1)}(\vec{k}-\vec{q}) = 2\omega_k\cdot(2\pi)^{n-1} |A_k|^2 \delta^{(n-1)}(\vec{k}-\vec{q}),
\eee
where we replaced terms like $A_q$ with $A_k$ since the $\delta$ function is zero wherever $\vec{k}\neq\vec{q}$. Thus, the normalized modes are given by 
\begin{align}
\varphi_{\vec{k}}(\vec{x},t) = \frac{1}{\sqrt{2\omega_k (2\pi)^{n-1}}} e^{i(\vec{k}\cdot\vec{x}-\omega_k t)}. \label{modes2} 
\end{align}
These, along with their complex conjugates, form a basis for the solution space to the Klein-Gordon equation.

As you are asked to verify in the exercises, plugging this into the $\phi$ expansion, and plugging in the commutation relations for the $\h{a}_{\vec{k}}$ results in the correct commutation relation between $\phi$ and $\pi$. 

Now that we've dealt with the explicit modes, we go back to the commutation relation (\ref{commutator}). It's the same ``great'' commutation relation we had for the oscillator (except that it's actually referring to an infinite number of operators, instead of just the two we had there). For the oscillator, we expressed the Hamiltonian in terms of $\h{x}$, and $\h{x}$ in terms of $\h{a}$. Here we express our Hamiltonian in terms of $\phi$, and $\phi$ in terms of all the $\h{a}_{\vec{k}}$. First, $H$ in terms of $\phi$ 
\be
H = \int d^{n-1}x \left(\frac{1}{2}\dot{\phi}^2 +\frac{1}{2} (\nabla \phi)^2 +\frac{1}{2}m^2\phi^2 \right). \nonumber
\ee
Then we plug in the expansion (\ref{expansion}) for $\phi$. It's a little tedious to calculate, but we just need to go term by term. We calculate the second term explicitly, and leave the others to you. Since $\nabla\varphi_{\vec{k}}=i\vec{k}\varphi_{\vec{k}}$, 
\begin{align*}
\int d^{n-1} x (\nabla\phi)^2 &= \int d^{n-1}x\int d^{n-1} k \int d^{n-1}q \left(i\vec{k}\h{a}_{\vec{k}}\varphi_{\vec{k}}-i\vec{k}\h{a}^{\D}_{\vec{k}}\varphi_{\vec{k}}^*\right)\cdot\left(i\vec{q}\h{a}_{\vec{q}}\varphi_{\vec{q}}-i\vec{q}\h{a}^{\D}_{\vec{q}}\varphi_{\vec{q}}^*\right). 
\end{align*}
And in this expression, we only calculate the first term arising from the product, since the others are almost identical
\begin{align*}
&\int d^{n-1}x\int d^{n-1} k \int d^{n-1}q \left(-\vec{k}\cdot\vec{q}\h{a}_{\vec{k}}\h{a}_{\vec{q}} \varphi_{\vec{k}}\varphi_{\vec{q}} \right) = \\ &\int d^{n-1}k\int d^{n-1} q \int d^{n-1}x \frac{-\vec{k}\cdot\vec{q}\h{a}_{\vec{k}}\h{a}_{\vec{q}}}{2(2\pi)^{n-1}\sqrt{\omega_k\omega_q}}e^{i\vec{x}\cdot(\vec{k}+\vec{q})} e^{-it(\omega_k+\omega_q)}= \\
 &\int d^{n-1}k\int d^{n-1} q \frac{-\vec{k}\cdot\vec{q}\h{a}_{\vec{k}}\h{a}_{\vec{q}}
 }{2\sqrt{\omega_k\omega_q}}e^{-it(\omega_k+\omega_q)}\delta^{(n-1)}(\vec{k}+\vec{q})= \int d^{n-1}k\frac{\vec{k}^2\h{a}_{\vec{k}}\h{a}_{-\vec{k}}}{2\omega_k}e^{-2i\omega_kt}. 
\end{align*}
Note that we integrated over $x$ first, which brought out the Delta function, which picked out $\vec{q}=-\vec{k}$ in the $q$ integral. Recall that $\omega_k^2=\vec{k}^2+m^2$. Thanks to this some terms cancel and some terms add up. The result is
\begin{align*}
H = \int d^{n-1} k \frac{\omega_k}{2}\left(\h{a}_{\vec{k}}\h{a}_{\vec{k}}^{\D}+\h{a}_{\vec{k}}^{\D}\h{a}_{\vec{k}}\right).
\end{align*}
Using (\ref{commutator}), this turns into 
\begin{align}
H = \int d^{n-1}k  \left(\h{n}_{\vec{k}}+\frac{1}{2}\delta^{(n-1)}(0)\right) \omega_k,  \label{qftH}
\end{align}
where $\h{n}_{\vec{k}}=\h{a}^{\D}_{\vec{k}}\h{a}_{\vec{k}}$. This is the $\vec{k}^{th}$ number operator which as we'll see, counts how many ``particles'' of momentum $\vec{k}$ there are. And yes, $H$ is infinite. There are two kinds of infinities here. First, the $\delta^{(n-1)}(0)$, which has to do with the fact that we used continuous plane wave solutions. This is not troublesome, since by simply using discrete solutions (using sums instead of integrals to define $\phi$), this infinity goes away.\footnote{In this case we would have $[\h{a}_{\vec{k}},\h{a}^{\D}_{\vec{q}}]=\delta_{\vec{q}\vec{k}}$, and 1 instead of $\delta^{(n-1)}(0)$.} This is equivalent to restricting our spacetime to be finite. 

The other infinity comes from the fact that we're summing over all $\vec{k}$, so the $\omega_k$ term will be infinite. This term is analogous to the zero-point (ground state) energy of the quantum harmonic oscillator, and it diverges because we assume our theory will work for arbitrarily large frequencies (or small wavelengths); clearly it won't. The way to deal with it is {\it renormalization}, which goes beyond our focus here. But conceptually, since we only care about \textit{differences} in energy, this constant doesn't affect observables. 

Now we can do the same thing we did for the harmonic oscillator. That is, first show that $[H,\h{a}_{\vec{k}}]=-\omega_k\h{a}_{\vec{k}}$. Then we consider an eigenstate $\ket{n_{\vec{k}}}$ of the Hamiltonian such that $H\ket{n_{\vec{k}}}=n_{\vec{k}}\omega_{k}\ket{n_{\vec{k}}}$, which corresponds to a state with $n_{\vec{k}}$ particles of momentum $\vec{k}$.\footnote{Just to be clear, we've ignored the factor of $\frac{1}{2}\delta^{(n-1)}(0)$, but it's not important here.} Just like for the oscillator, we have $H(\h{a}_{\vec{k}}\ket{n_{\vec{k}}}) = \omega_k(n_{\vec{k}}-1)(\h{a}_{\vec{k}}\ket{n_{\vec{k}}})$. This means that $\h{a}_{\vec{k}}\ket{n_{\vec{k}}}$ is an eigenstate of $H$ with energy corresponding to a state with $n_{\vec{k}}-1$ particles. In other words, the operator $\h{a}_{\vec{k}}$ \textit{annihilated} a particle of momentum $\vec{k}$. For this reason, we call $\h{a}_{\vec{k}}$ the annihilation operator.
On the other hand, $\h{a}^{\D}_{\vec{k}}$ is the \textit{creation} operator; it creates a particle of momentum $\vec{k}$ (as you may verify). 

The Hamiltonian is a positive definite operator. Then, it has a ground state, a state with the lowest eigenvalue that we name $\ket{0}$ such that $H\ket{0}=E_{gs}\ket{0}$ (technically, $E_{gs}$ is infinity but fixed; we're only focusing on the $\h{n}_{\vec{k}}$ part of the Hamiltonian). By definition, this is a state such that $\h{a}_{\vec{k}}\ket{0}=0$ for all $\vec{k}$ (otherwise it would not be the ground state). This state is called the \textit{Fock vacuum}, a state with no particles
\be
\braket{0|\h{n}_{\vec{k}}|0}=0,\quad\quad \forall\ \vec{k}.
\ee
 From the Fock vacuum we can construct the whole Hilbert space of states, also called \textit{Fock space}. The way to do this is by acting on the vacuum with the creation operators. 

Above is the promised physical relevance of the operators $\h{a}_{\vec{k}}$ and $\h{a}_{\vec{k}}^{\D}$. But recall, these operators appeared simply as the coefficients of positive and negative frequency modes in our expansion of $\phi$! That's all they are. We did some calculations, however, that showed us we can interpret some physics behind them. They create and annihilate particles, and they also helped us define our Fock space, and Fock vacuum. 

To summarize, the way we define the vacuum depends on the way we separate our space of classical field solutions into positive and negative frequency modes. In the next section we'll see that this is not so easy in curved spacetime, and that even in flat spacetime there are many inequivalent ways to do the separation. 


\subsection{Scalar Fields in Curved Spacetime}

We begin our study of scalar fields in  curved spacetime by addressing an awkward point that you may have noticed. We were working in a theory that supposedly incorporates special relativity (and thus should present Lorentz-invariant results), and yet we have expressions that seem very dependent on our choice of reference frame. For example, to characterize positive frequency modes, we said that $\pd_t \varphi_{\vec{k}}=-i\omega_k\varphi_{\vec{k}}$, which uses specific components of the coordinates $x^{\mu}$ and the vector $k^{\mu}$ ($x^0=t$ and $k^0=\omega_k$). 

We can show that positive frequency is a Lorentz invariant definition. We do $x^{\mu}\to x^{\mu'}=\tensor{\Lambda}{^{\mu'}_\mu}x^{\mu}$, where $\tensor{\Lambda}{^{\mu'}_\mu}$ is any Lorentz transformation. First we note that $\varphi_{\vec{k}}$ doesn't change, since it is equal to $Ae^{ik_{\mu}x^{\mu}}$, and $k_{\mu}x^{\mu}$ is an invariant quantity.\footnote{Explicitly, $k'_{\mu}x'^{\mu}=\Lambda\indices{_\mu^\alpha}k_{\alpha}\Lambda\indices{^\mu_\beta}x^{\beta}=k_{\alpha}\Lambda\indices{_\mu^\alpha}\Lambda\indices{^\mu_\beta}x^{\beta}=k_{\mu}x^{\mu}$ because by raising an index to $\Lambda\indices{^\nu_\alpha}\eta_{\nu\mu}\Lambda\indices{^\mu_\beta}=\eta_{\alpha\beta}$ we obtain the Kronecker delta.}
 Additionally, $\partial_t\to \tensor{\Lambda}{^{\mu}_{t'}}\partial_{\mu}\equiv \tensor{\Lambda}{^{\mu}_0}\pd_{\mu}$ under this transformation, so we have 
\begin{align*}
\partial_{t'}\varphi_{\vec{k}'}=\tensor{\Lambda}{^{\mu}_{0}}\partial_{\mu}Ae^{ik_{\nu}x^{\nu}} =  \tensor{\Lambda}{^{\mu}_{0}}\delta^{\nu}_{\mu}ik_{\nu}\varphi_{\vec{k}'}=i\tensor{\Lambda}{^{\nu}_{0}}k_{\nu}\varphi_{\vec{k}'}=-i\omega_k'\varphi_{\vec{k'}}, 
\end{align*}
where $\omega_k'=-\tensor{\Lambda}{^{\nu}_{0}}k_{\nu}$. Which, as we show in a moment, tells us that indeed our definition of a positive frequency mode is frame-invariant. And why should we care so much about positive frequency modes? Recall from the previous section that the concept of vacuum is constructed from the \textit{coefficients} of positive frequency modes. Therefore, since positive frequency modes are Lorentz-invariant, so is the notion of a vacuum $\ket{0}$. Can we make this more general? Is our definition of positive frequency modes invariant under an \textit{arbitrary} change of coordinates? We'll see a concrete example of this later, when we discuss Bogoliubov coefficients. 

Now we check what happens when we have a Lorentz transformation, rather than an arbitrary one, as in the argument above. For $\varphi_{\vec{k}'}$ to be a positive frequency mode with respect to $t'$, the quantity $\omega'_k=- \tensor{\Lambda}{^\mu_0}k_{\mu}$ must be positive. We will prove now that this is guaranteed if $\omega_k$ is positive and $\tensor{\Lambda}{^{\mu'}_\mu}$ is an  orthochronous Lorentz transformation.\footnote{An orthochronous Lorentz transformation does not change the direction of time, which can be expressed as the positivity of $\tensor{\Lambda}{^0_0}>0$.} 


We split the quantity $-\tensor{\Lambda}{^\mu_0}k_{\mu} = -\Lambda_{\mu0}k^{\mu} = -\Lambda_{00}\omega_k -\Lambda_{i0}k^i $. We know that $\omega_k=k^0$ is positive and $\Lambda_{00}<0$,\footnote{In detail: $\Lambda_{00}=\eta_{\mu0}\tensor{\Lambda}{^{\mu}_0}=- \tensor{\Lambda}{^{0}_0}<0$ because the transformation is orthochronous!} so $-\Lambda_{00}\omega_k$ is positive always, and we are on the right track. On the other hand $-\Lambda_{i0}k^i$ does not necessarily have to be positive; we want to show that even when it's negative, the $-\Lambda_{00} \omega_k$ term ``wins''. We will show that the $\Lambda_{00}$ part ``beats'' the $\Lambda_{i0}$ part, and likewise for the $\omega_k$ versus the $k^i$.

To show it, recall that a Lorentz transformation $\Lambda$ solves the matrix equation $\Lambda^T\eta \Lambda=\eta$, which in components is written as  $\tensor{\Lambda}{^\mu_\alpha} \eta_{\mu\nu} \tensor{\Lambda}{^\nu_\beta} = \eta_{\alpha\beta}$. If we look at the $00$ component we have 
\bee
\tensor{\Lambda}{^0_0} \eta_{00} \tensor{\Lambda}{^0_0}+\tensor{\Lambda}{^i_0} \eta_{ij} \tensor{\Lambda}{^j_0} =-(\Lambda_{00})^2 +\sum_{i=1}^{n-1} (\Lambda_{i0})^2=-1,
\eee
or $(\Lambda_{00})^2 = 1+ \sum_i (\Lambda_{i0})^2$. Having established this property, we look at $\omega_k$. We know that the condition $\omega_k$ must satisfy is $\omega_k^2=\vec{k}^2+m^2=k_ik^i+m^2$. Then
\begin{align*}
(\Lambda_{00})^2\omega_k^2 &= \left(1+ \sum_{i=1}^{n-1} (\Lambda_{i0})^2\right) \left(\sum_{j=1}^{n-1} (k^j)^2 +m^2\right)  >  \sum_{i=1}^{n-1} \sum_{j=1}^{n-1} (\Lambda_{i0})^2(k^j)^2 \geq   \left(\sum_{i=1}^{n-1} \Lambda_{i0}k^i\right)^2,
\end{align*}
where first we simply subtracted $1$ and $m^2$ to get a strict inequality and then we made use of the the Cauchy-Schwarz inequality. 
Now we can conclude that $|\Lambda_{00}\omega_k| > |\Lambda_{i0}k^i|$. And since $-\Lambda_{00}\omega_k$ is positive, $-\Lambda_{00}\omega_k-\Lambda_{i0}k^i>0$ always. Then, $\omega_k'> 0$. 

Therefore, an arbitrary (orthochronous) Lorentz transformation leaves the notion of positive frequency modes invariant. It was critical to assume that it was a Lorentz transformation, we suggest you to try to find a coordinate transformation that does not leave positive frequency modes invariant, there are many! 

So we know that the notion of positive frequency modes will not be invariant under an arbitrary transformation, then in general two different observers with reference frames that are not related by a Lorentz transformation will have different notions of positive frequency. This happens in curved spacetimes where we cannot define inertial frames globally, but it also happens for flat spacetimes where non inertial frames and observers may be considered.

Now we repeat for curved spacetimes the procedure we've carried out for the scalar field in flat spacetime to compare the situations. That is, we quantize a scalar field in a fixed curved spacetime in the usual manner, and see how different notions (like the Fock vacuum) change.

The usual changes to curved spacetime must be made (i.e. $\eta_{\mu\nu}\to g_{\mu\nu}$, $\pd_{\mu}\to\nabla_{\mu}$). The action for a scalar field also changes. First,  the integral measure  picks up a factor of $\sqrt{-g}$, where $g$ is the determinant of the metric (as is usual for integrating over a manifold). So the action is
\be
S = \int d^nx \sqrt{-g}  \left(-\frac{1}{2}g^{\mu\nu}\partial_{\mu}\phi \partial_{\nu}\phi - \frac{1}{2}m^2\phi^2 \right). 
\ee 
Additionally, the partial derivatives should be covariant derivatives, but since we have a scalar field it doesn't matter. 

Another change is a possible coupling of the scalar field to the Ricci scalar $R$, which may describe an interaction of the field with the background spacetime curvature. An interaction typically considered is of the form $\frac{1}{2}\xi R\phi^2$, where $\xi$  is a coupling constant. To take it into account, it must simply be added in the action's integral. 

The equation of motion (taking into account the interaction) changes a little now
\be
(\square -m^2-\xi R)\phi=0, \label{motion3}
\ee
where $\square$ now means $\nabla_{\mu}\nabla^{\mu}$. 


The Lagrangian density that produces the previous equation is $\Lc=\sqrt{-g} (-\frac{1}{2}g^{\mu\nu}\pd_{\mu}\phi\pd_{\nu}\phi-\frac{1}{2}m^2\phi^2-\frac{1}{2}\xi R\phi^2)$, so the conjugate momentum is 
\bee
\pi = \frac{\pd \Lc}{\pd \dot{\phi}}=-\sqrt{-g} g^{\mu 0}\pd_{\mu}\phi.
\eee 
The conjugate momentum is defined at spacelike ``constant time'' hypersurfaces. 
To quantize the theory we promote $\phi$ and $\pi$ to operators, and impose a new set of commutation relations 
\be
[\phi(\vec{x}),\pi(\vec{y})]= \frac{i}{\sqrt{-g}} \delta^{(n-1)}(\vec{x}-\vec{y}). 
\ee
Just like for the Klein-Gordon equation, we can find an orthonormal basis of solutions to (\ref{motion3}) but now they won't be so useful, as we'll see. First we generalize the inner product on the solution space for curved background spacetimes
\be
\braket{\phi_1|\phi_2} = i\int_{\Sigma} d^{n-1}x \sqrt{\gamma} n^{\mu}(\phi_1^*\nabla_{\mu}\phi_2-\phi_2\nabla_{\mu}\phi_1^*), \label{inner3}
\ee
which is taken over a spacelike hypersurface $\Sigma$ with normal $n^{\mu}$ and induced metric $\gamma_{ij}$ ($\gamma\equiv \det \gamma_{ij}$). You can verify that this reduces to  (\ref{inner2}) in Minkowski spacetime. This inner product is independent of the hypersurface we take (that is, it doesn't depend on which ``time-slice'' we take), thanks to the divergence theorem and because $\phi_{1,2}$ are solutions of the equation of motion (see the flat spacetime proof in (\ref{stokes})). 

With an adequate inner product defined, let's say we found an orthonormal set of modes $f_i$ (and their conjugates $f_i^*$) to expand a general solution  
\bee
\phi = \sum_i\left( \h{a}_if_i+\h{a}^{\D}_i f_i^*\right), 
\eee
where we now use a discrete set of modes rather than a continuous one, for simplicity. 
And $\pi$ is 
\begin{align*}
\pi =- \sqrt{-g}g^{\mu0} \sum_i \left(\h{a}_i\pd_{\mu}f_i + \h{a}_i^{\D}\pd_{\mu}f_i^*\right).
\end{align*}

The coefficients $\h{a}_i$ will satisfy the usual commutation relations; we leave it to you to show this. The steps are analogous to what was done with the harmonic oscillator, and scalar fields in flat spacetime (i.e. use the inner product formalism, with the assumption that an orthonormal basis separated into positive and negative frequency modes exists). 




So we have 
\be
[\h{a}_i,\h{a}_j^{\D}] = \delta_{ij} \nonumber,
\ee
and
\be
[\h{a}_i,\h{a}_j] =[\h{a}_i^{\D},\h{a}_j^{\D}] =0.
\ee
Once again, $\h{a}_i$ and $\h{a}_j^{\D}$ are annihilation and creation operators, and there is a vacuum state $\ket{0_A}$ defined by the fact that $\h{a}_i\ket{0_A}=0$ for all $i$. 



This all seems identical to what we had before in flat spacetime, so what's new?

 Let us consider another set of orthonormal modes to expand  $\phi$, call them $g_i$. Then
\begin{align*}
\phi = \sum_i \left(\h{b}_i g_i + \h{b}_i^{\D}g_i^*\right),
\end{align*}
where $\h{b}_i$ are coefficients in the expansion, different from the $\h{a}_i$. 
Just like before, they'll satisfy commutation relations $[\h{b}_i,\h{b}^{\D}_j]=\delta_{ij}$, $[\h{b}_i,\h{b}_j]=[\h{b}_i^{\D},\h{b}_j^{\D}]=0$. We can use them to define a Fock space and vacuum, such that $\h{b}_i\ket{0_B}=0$. But is this vacuum the same as $\ket{0_A}$? We now show that in general this is not the case. 

First, we carry out a Bogoliubov transformation: this is simply to expand the modes $f_i$ in terms of the $g_i$, and vice-versa. It's possible since both sets form a complete basis. 
\begin{align*}
g_i &= \sum_j\left( \alpha_{ij}f_j+\beta_{ij}f^*_j\right), \\
f_i &= \sum_j \left(\gamma_{ij}g_j+\lambda_{ij}g_j^*\right).
\end{align*}
We can find relations between all these coefficients. Recall that the modes are properly normalized, so that $\braket{f_i|f_j}=\delta_{ij}$, $\braket{f_i^*|f_j^*}=-\delta_{ij}$, and similarly for the $g$'s. So $\alpha_{ij}= \braket{g_i|f_j}$. On the other hand, $\gamma_{ij}=\braket{f_i|g_j}$, which implies that $\gamma_{ij}=\alpha_{ji}^*$. You may check this in more detail, and also (analogously) note that $\lambda_{ij}=-\beta_{ji}$. So have
\begin{align}
g_i &= \sum_j \left(\alpha_{ij}f_j+\beta_{ij}f^*_j\right),\nonumber \\
f_i &= \sum_j\left( \alpha^*_{ji}g_j-\beta_{ji}g_j^*\right).\label{bog} 
\end{align}
These transformations imply similar transformations between the creation/annihilation operators corresponding to each set of modes. This is because 
\bee
\phi = \sum_i \left( \h{a}_if_i+\h{a}_i^{\D}f_i^* \right)= \sum_i \left(\h{b}_ig_i+\h{b}_i^{\D}g_i^*\right). 
\eee
If we plug in our Bogoliubov transformations for one side, after matching the coefficients of each mode (we leave this to you) we arrive at 
\begin{align}
\h{a}_i &= \sum_j \left(\alpha_{ji}\h{b}_j+\beta_{ji}^*\h{b}_j^{\D}\right), \nonumber \\ 
\h{b}_i &= \sum_j \left(\alpha_{ij}^*\h{a}_j-\beta_{ij}^*\h{a}_j^{\D}\right). 
\end{align}
Here is the key issue when we take quantum field theory from flat spacetime to curved spacetime. Our annihilation operators for one observer (e.g. $\h{a}$) are not the annihilation operators of the other observer. Rather, they mix both creation and annihilation operators. This will result (in general) in one observer's vacuum having particles for another observer. Let's verify this explicitly. If we consider the vacuum $\ket{0_A}$ of observer $\mathcal{A}$ who uses the expansion with the $f_i$, we have $\h{a}_i\ket{0_A}=0$ for all $i$. The expected number of particles of type $i$ for $\mathcal{A}$ will thus be $\braket{\h{n}_{a_i}}=\braket{0_A|\h{a}^{\D}_i\h{a}_i|0_A}=0$. But according to $\mathcal{B}$, the number of particles in the state $\ket{0_A}$ is 
\begin{align*}
\braket{\h{n}_{b_i}} &= \braket{0_A|\h{b}^{\D}_i\h{b}_i|0_A} \\
&= \Braket{0_A|  \left(\sum_j \alpha_{ij}\h{a}_j^{\D}-\beta_{ij}\h{a}_j \right) \left(\sum_j \alpha_{ij}^*\h{a}_j-\beta_{ij}^*\h{a}_j^{\D}\right)|0_A}\\
& = \Braket{0_A| \sum_j\sum_k \beta_{ij}\beta_{ik}^*\h{a}_j\h{a}_k^{\D}|0_A} \\
& = \sum_j\sum_k \beta_{ij}\beta_{ik}^* \Braket{0_A|\left( \h{a}^{\D}_k \h{a}_j+\delta_{jk}\right) |0_A} \\
&= \sum_j |\beta_{ij}|^2,
\end{align*}
where in going from the second line to the third, we used that any terms with $\h{a}_j$ on the right annihilate $\ket{0_A}$, while any terms with $\h{a}_j^{\D}$ to the left annihilate $\bra{0_A}$. To go from the third line to the fourth we used $[\h{a}_j,\h{a}^{\D}_k]=\delta_{jk}$. The result is that $\mathcal{A}$'s vacuum isn't a vacuum for $\mathcal{B}$! In the $\beta$ coefficients there is information regarding the amount of particles that observer $\mathcal{B}$ sees in $\mathcal{A}$'s vacuum. This makes sense if we look at (\ref{bog}), since the $\beta$ coefficients show the mixing between positive and negative frequency modes of the different observers. 

Note that the above mixing of positive and negative frequency modes does not occur if the modes correspond to two inertial observers, i.e. two observers related by a Lorentz transformation. This is what we showed at the beginning of this section, which in this new language translates into saying that $\beta_{ij}=0$ for all $i,j$. Once again, the vacuum concept is invariant under Lorentz transformations. The problem arises when two reference frames are related by more general transformations. 

Now, you might wonder: If the vacuum for an observer has particles for another observer, is our interpretation of what a particle is (in quantum field theory) correct? What is actually meant when we say that an observer detects particles? As Unruh showed \cite{Unruh:1976db}, particle detectors use their proper times to define positive and negative frequency modes (and thus ``particles''). One model he used for a particle detector  is a Schr$\ddot{\text{o}}$dinger particle in a box being accelerated. If the particle starts out in its ground state (the vacuum), as the box accelerates there's a non-zero probability that the particle goes into an excited state. This is a non-ambiguous example of observing a ``particle'', and it's a specific instance of the Unruh effect. We'll look at it in the next section, and argue why in general an accelerating observer detects particles in the vacuum of an inertial observer. 

To finish up here, we consider in what cases we can actually find positive and negative frequency modes all over spacetime in order to define Fock vacua and related concepts. Roughly, we need to be able to separate our wave equation (e.g. $(\square-m^2-\xi R) \phi=0$) globally into a ``space'' part and a ``time'' part. This amounts to expressing $\square=\nabla_{\mu}\nabla^{\mu}$ as $-\pd_t^2+\mathcal{D}$, where $\mathcal{D}$ is some differential operator independent of time. In this case we can find separable solutions $f_{\omega}(t,\vec{x})=e^{-i\omega t} g_{\omega}(\vec{x})$, such that $\pd_t f_{\omega} = -i\omega f_{\omega}$ and we can have a clear separation of positive and negative frequency modes. 

The ability to carry out this separation globally rests on the possibility of defining a hypersurface orthogonal to a timelike Killing vector $K^{\mu}$ all over spacetime. However, if we allow local notions of time the separation is not unique and therefore different positive frequency notions can be used, which is what results in different observers defining different vacua. 

\section{Unruh Effect}
\label{unruheffect}

The Unruh effect clearly illustrates the last concepts discussed. A vacuum state for one observer is not necessarily a vacuum state for another observer. In other words, different observers can use different positive frequency modes, and will therefore see different vacua. 

In the following we will study and compare the quantization of a field by an inertial observer with the quantization of a field by an accelerated observer. Both observers will be living in the same flat spacetime. For simplicity we consider a two dimensional spacetime, therefore, it is a one dimensional world with a time dimension. And as a probe field we take a free massless scalar field $\phi$. This simplification still captures the phenomenon while the generalization to real four dimensional spacetime and other fields is straightforward.

The result is that the quantization won't be equivalent. The physical consequence is that the accelerated observer measures a temperature proportional to their acceleration. This is the Unruh effect. Now we go into the details that allow us to reach this conclusion.

\subsection{Rindler Spacetime} 

\begin{figure}[ht!]
\centering
\includegraphics[scale=0.3]{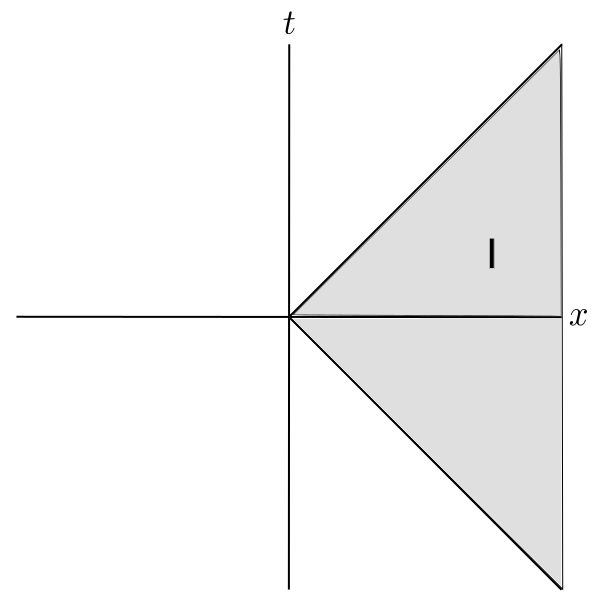}\caption{Rindler spacetime}\label{rindler2}
\end{figure}

The first thing we must do is describe the path that a uniformly accelerating observer follows. Consider a flat spacetime described by coordinates $t$ and $x$; and line element $ds^2=-dt^2+dx^2$. What does uniform acceleration mean? Recall that if $\alpha$ is the norm of the proper acceleration we have $\alpha\equiv\sqrt{a_{\mu}a^{\mu}}$, where $a^{\mu}=d^2x^{\mu}/d\tau^2$ (with $\tau$ the proper time along the trajectory).  We want this to be constant. Let the path of the accelerated observer be given by some functions $t(\tau)$ and $x(\tau)$. Then we have 
\bee
\alpha^2 = -\left(\frac{d^2t}{d\tau^2}\right)^2+\left(\frac{d^2x}{d\tau^2} \right)^2.
\eee
The path of the accelerating observer, as you may verify, will be described by
\begin{align}
t(\tau)= \frac{1}{\alpha}\sinh(\alpha\tau), \hspace{1cm} x(\tau)=\frac{1}{\alpha}\cosh(\alpha\tau). 
\end{align}
This parametrizes the curve given by $x^2=t^2+1/\alpha^2$, which corresponds to a hyperboloid with asymptotes given by null paths $x=\pm t$. It's important to note that the accelerating observer is always in the $x>|t|$ region (we call it region I, see Fig. \ref{rindler2}, and you can verify this statement). In fact, the observer comes from $x=+\infty$ with almost the speed of light and decreases his speed as he approaches the origin. Then, in the closest point to the origin he has zero speed, and then, increases his speed again and goes towards $x=+\infty$ reaching almost the speed of light. Thus a Rindler observer {\it bounces} close to the origin between two asymptotic situations with almost the speed of light.

Now we want to define a new set of coordinates $(\eta,\xi)$ that will be more naturally adjusted to accelerated observers, and cover region I.

This region I is  the background in which the Unruh effect takes place. It is basically flat spacetime, but we call it Rindler spacetime since it only consists of a wedge of flat spacetime (which changes things quite a lot). Additionally, we describe Rindler spacetime using a seemingly strange coordinate system. However, it's not strange, it's simply a coordinate system related to accelerated observers.

The way to construct this coordinate system, intuitively, is to consider the whole set of possible constantly accelerating observers. For the path of a given accelerated observer, we'll say that the parameter $\xi$ is constant, and only $\eta$ changes. We can think of $\eta$ as a time parameter along the trajectory of the observer. Meanwhile, to cover all of region I we need more than just one observer---we need an infinite amount! $\xi$ will parametrize the continuous ``switching'' between observers of different accelerations (each corresponding to different hyperboloids in region I). So some useful coordinates could be defined (implicitly) by 
\begin{align}
t(\eta,\xi)=\frac{1}{a}f(\xi)\sinh(a\eta), \hspace{1cm} x(\eta,\xi)=\frac{1}{a}f(\xi)\cosh(a\eta). \nonumber
\end{align}

\begin{figure}[ht!]
\centering
\includegraphics[scale=0.2]{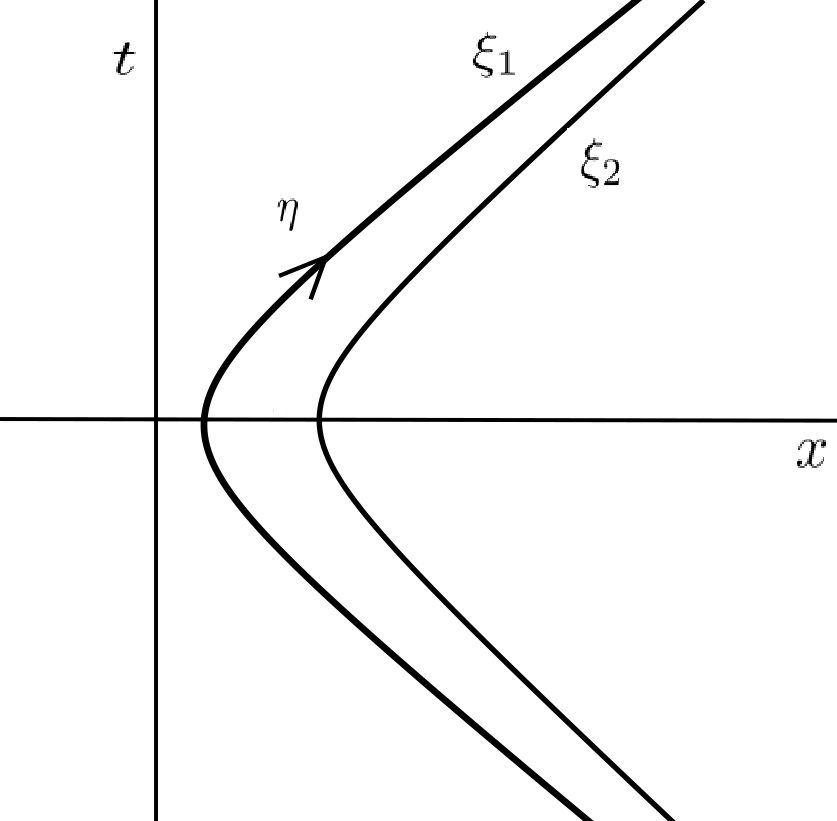}\caption{Using accelerated observers to construct a coordinate system}\label{rindler1}
\end{figure}

In figure \ref{rindler1}, we sketched out the trajectories of two accelerated observers, each labeled by a constant value of $\xi$ and sweeping out all values of $\eta$.

Next, note that these coordinates satisfy the following
\bee
x^2-t^2 = \frac{f(\xi)^2}{a^2}.
\eee
Comparing the above to what we had found for an observer with acceleration $\alpha$, we see that an observer corresponding to the coordinate $\xi_0$ has acceleration $a/f(\xi_0)$. 
As you are asked to sketch out in the exercises, we have that an observer with higher acceleration gets closer to the origin. That is, for lower $f(\xi)$ we are closer to the origin. This motivates us to require that $f$ be an increasing function. As $\xi$ increases, the accelerated observer along that $\xi$ is farther from the origin. 

Considering this last requirement, we choose $f(\xi) = \exp(a\xi)$. This is by no means a unique choice.\footnote{Another commonly used option is $f(\xi)=\xi$.} Then, the coordinate transformation is 
\begin{align}
t = \frac{1}{a}e^{a\xi}\sinh(a\eta), \hspace{1cm} x = \frac{1}{a}e^{a\xi}\cosh(a\eta).
\end{align}
One detail we've forgotten is: What are the ranges of $\xi$ and $\eta$? We have simply $\xi,\eta\in\mathbb{R}$. This results in $t\in\mathbb{R}$ and $x\in(t,\infty)$, which is precisely what we need to cover region I.\footnote{If we had instead used $f(\xi)=\xi$, the range of $\xi$ would be $(0,\infty)$.}

Now to introduce the terminology: \textit{Rindler wedge} is simply region I of Minkowski spacetime usually described by the coordinates we have defined. A \textit{Rindler observer}, meanwhile, is an observer moving along constant $\xi$ (which means stationary in our Rindler coordinates). But recall that from the point of view of flat spacetime, a Rindler observer is just someone with uniform acceleration.

\subsection{Positive Frequency Modes in Rindler Spacetime} 

The coordinate $\eta$ seems to be like a time coordinate for Rindler spacetime, so let's see if $\pd_{\eta}$ is a timelike Killing vector. Using the chain rule, 
\begin{align*}
\pd_{\eta} = \frac{\pd x}{\pd\eta}\pd_x + \frac{\pd t}{\pd\eta}\pd_t = e^{a\xi}\sinh(a\eta)\pd_x+e^{a\xi}\cosh(a\eta)\pd_t = a(t\pd_x+x\pd_t).
\end{align*}
To show that $\pd_{\eta}$ is timelike, we must show that $(\pd_{\eta})^{\mu}(\pd_{\eta})_{\mu}<0$ (where $\mu$ is a dummy index to be summed over, and $\eta$ is a fixed coordinate). Indeed,
\bee
(\pd_{\eta})^{\mu}(\pd_{\eta})_{\mu} = g_{\mu\nu}(\pd_{\eta})^{\mu}(\pd_{\eta})^{\nu}=-(\pd_{\eta}^t)^2+(\pd_{\eta}^x)^2=a^2(-x^2+t^2)<0,
\eee
since Rindler spacetime is Minkowski spacetime restricted to the region $x>|t|$. It is similarly straightforward to verify that this is a Killing vector (using Killing's equation $\nabla_{(\mu}K_{\nu)}=0$ for a Killing vector field $K$).

\begin{figure}[ht!]
\centering 
\includegraphics[scale=0.4]{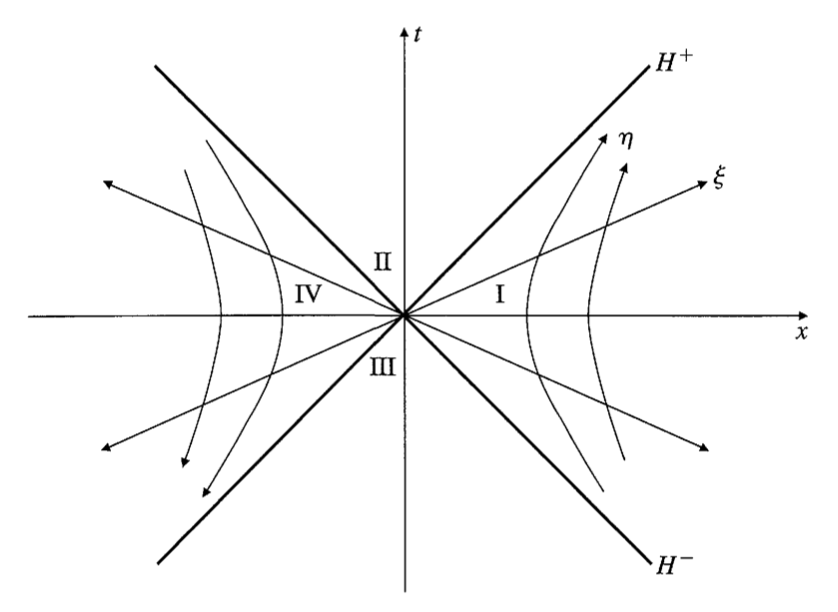}\caption{Regions of Minkowski spacetime (taken from \textit{Spacetime and Geometry}, Carroll)} \label{rindler3} 
\end{figure}

Now, since $\pd_{\eta}$ is a timelike Killing vector in region I, we can use it to define positive and negative frequency modes (i.e. $f$ is a positive frequency mode if $\pd_{\eta}f=-i\omega f$ with $\omega>0$). However, in the other regions of Minkowski spacetime, it won't be so simple. Figure \ref{rindler3} labels the regions we'll separate Minkowski spacetime into. Rindler spacetime is region I, and we'll be interested in \textit{extending} it to the other regions later on. 

We can now find positive frequency modes in Rindler spacetime using $\pd_{\eta}$. We leave it to you to verify that the metric using the coordinates $(\eta,\xi)$ is $ds^2= e^{2a\xi}(-d\eta^2+d\xi^2)$. In these coordinates the Klein-Gordon equation is 
\begin{align}
\square\phi = e^{-2a\xi}(-\pd_{\eta}^2+\pd_{\xi}^2)\phi =0.
\end{align}
This is of the same form as (\ref{KG}) (but with $m=0$), which means that the solutions will also be of the same form. That is, we can expand $\phi$ in the following positive frequency modes (and their conjugates)
\begin{align}
f_{k} = \sqrt{\frac{1}{4\pi\omega_k}}e^{-i\omega_k\eta+ik\xi}.
\end{align}
Here $\omega_k=|k|$ since we're in two dimensions. Let's see that these modes are properly normalized using (\ref{inner3}): we will integrate over a constant $\eta$ surface. Any normal vector $n^{\mu}$ will only have a component along $\pd_{\eta}$, and if it is to be a unit vector, it will satisfy $f_{\mu\nu}n^{\mu}n^{\nu}=-e^{2a\xi}(n^{\eta})^2=-1$. Thus, $n^{\eta}=e^{-a\xi}$. The determinant of the induced metric will be $\gamma=e^{2a\xi}$, so $\sqrt{|\gamma|}=e^{a\xi}$. Then in the integral the $n^{\mu}\sqrt{|\gamma|}$ term reduces to $\delta^{\mu\eta}$. So 
\begin{align*}
\braket{f_{k}|f_{q}} &= i \int (f_{q}^*\pd_{\eta}f_{k}-f_{k}\pd_{\eta}f_{q}^*) d\xi \\
&= \int \frac{1}{4\pi\sqrt{\omega_k\omega_q}} (\omega_k+\omega_q)\left(e^{i(\omega_q-\omega_k)\eta-i(q-k)\xi}\right)d\xi \\
&= \delta(q-k) .
\end{align*}
Form the previous equation it is easy to check that the negative frequency modes satisfy $\braket{f^*_k|f^*_q}=-\delta(k-q)$. We leave it to you to verify that the positive frequency and negative frequency modes are indeed orthogonal. 



\subsection{Extension of Rindler Modes}  \label{elegant}

The positive frequency modes we defined above are only positive frequency in Rindler spacetime, region I. They can be used to define the quantization of the field for an accelerated observer. 


Now, we ask how the vacuum state of an inertial observer is seen by an accelerated observer. This requires the calculation of the Bogoliubov coefficients that relate the Rindler and the Minkowski modes.  In particular we can compute the expectation value of the Rindler number of particles operator in the Minkowski vacuum state $\ket{0_M}$. 

The direct computation is straightforward using complex analysis integration techniques, but it may be made easier by using a different representation of the modes in the Minkowski spacetime. A representation that shares the same Minkowski vacuum state but is built using {\it analytic extensions} of the Rindler modes to the whole spacetime (not just region I).   

These modes will be different from the $\varphi_{\vec{k}}$ natural for the inertial observers, but we'll provide an elegant argument (Unruh's idea) to show that they characterize the same vacuum state $\ket{0_M}$. So essentially we have Rindler modes $f_k$, and we'll construct another set of modes $h_k$; these will be an extension inspired by the $f_k$, so that at the end comparing the two sets of modes will be straightforward.

First, we define the positive frequency modes analogous to the $f_k$ in region IV (see figure \ref{rindler3}). We have to define a new set of coordinates to cover this, since ($\eta,\xi$), taking all possible real values, only cover region I. We use 
\begin{align}
t = -\frac{1}{a}e^{a\delta}\sinh(a\gamma), \hspace{1cm} x = -\frac{1}{a}e^{a\delta}\cosh(a\gamma), 
\end{align}
where $\gamma$ and $\delta$ are both in $(-\infty,\infty)$. 
Note that $\pd_{\gamma}$ is timelike here, but it is past-directed (in the sense that as $\gamma$ increases, $t$ decreases). So we use $\pd_{-\gamma}=-\pd_{\gamma}$ to define positive frequency modes in region IV, instead of $\pd_{\gamma}$. Then our positive frequency modes here are
\begin{align}
g_k = \frac{1}{\sqrt{4\pi\omega_k}}e^{i(\omega_k\gamma+k\delta)}.
\end{align}
You may check that this is indeed positive frequency in region IV. These $g_k$ modes are equivalent to $f_k$ modes but for observers accelerating towards the left in a symmetrical fashion than the accelerating observers defining the Rindler wedge I.

Ok, now we've defined positive frequency modes in regions I and IV, $f_k$ and $g_k$. They are functions that are only defined in their respective regions but we want them as seeds to build a new basis in the whole spacetime. Then, we'll define them so that they cover both regions I and IV. The way to do this is simple, define the following functions
\begin{align}
f_k^{(1)} = \begin{cases} 
\frac{1}{\sqrt{4\pi\omega_k}}e^{-i(\omega_k\eta-k\xi)}, \hspace{0.5cm} &\text{I}\\
0,  &\text{IV}
\end{cases} 
\end{align}
That is, if we're in region I, the mode $f_k^{(1)}$ is simply the positive frequency mode we defined earlier for region I, while in region IV it is 0. Similarly for $g_k$ we have
\begin{align}
f_k^{(2)} = \begin{cases} 
0, &\text{I}\\
 \frac{1}{\sqrt{4\pi\omega_k}}e^{i(\omega_k\gamma+k\delta)}, \hspace{0.5cm}  &\text{IV}
\end{cases} 
\end{align}
Each function is positive frequency in the region where it is not zero, with respect to the timelike Killing vectors $\partial_\eta$ and $-\partial_\gamma$, respectively.  

Now, to have a complete basis for scalar functions in Minkowski spacetime we need to define them over a Cauchy surface. We can choose the $t=0$ line or any straight line crossing the origin while lying inside region I and IV. However, as we make explicit in a moment, if we use directly the functions $f_k^{(1)}$ and $f_k^{(2)}$ over a line they do not fit smoothly at the origin, for instance, the function $f_k^{(1)}+f_k^{(2)}$ is not analytic at the origin $x=t=0$.


Besides their non-analiticity at the origin, the set of modes $f_k^{(1)}$ and $f_k^{(2)}$ are not suitable to be compared with the Minkowski modes. The reason is that with respect to the Killing vector $\pd_t$, defining time for the inertial observer, they are not positive frequency modes. Using  $f_k^{(1)}$ and $f_k^{(2)}$ as seeds we would like to have positive frequency modes with respect to Minkowski time. This is so that the vacuum state condition can be expressed with annihilation operator associated to both the usual Minkowski modes, $\varphi_{\bf k}$, and the new Rindler inspired set of modes (denoted by $h_k$ in the following).

To build analytic modes through the origin we'll use a trick. We use the relation of the Rindler coordinates, that partially cover the spacetime, with the Minkowski coordinates, that cover the whole spacetime. This will allow us to extend our Rindler modes of each wedge to the remaining regions. This is called {\it analytic extension}; we use the analyticity property of complex functions to by-pass the problem at the origin and define well-behaved functions.

First, we need the relation between Rindler coordinates and Minkowski coordinates in different regions. In region I
\bee
a(-t+x)=a\left(-\frac{1}{a}e^{a\xi}\sinh(a\eta)+\frac{1}{a}e^{a\xi}\cosh(a\eta)\right)=e^{-a(\eta-\xi)},
\eee
while in region IV we have 
\bee
a(-t+x)= a\left(\frac{1}{a}e^{a\delta}\sinh(a\gamma)-\frac{1}{a}e^{a\delta}\cosh(a\gamma)\right)=-e^{-a(\gamma-\delta)}.
\eee
So we can say that in region I, when $k>0$ (so $\omega_k=k$)
\be
\sqrt{4\pi\omega_k}f_k^{(1)}=e^{-i\omega_k(\eta-\xi)}=(e^{-a(\eta-\xi)})^{i\omega_k/a}=a^{i\omega_k/a}(-t+x)^{i\omega_k/a}.
\label{funI}
\ee
If we let $(t,x)$ to be outside region I we may define an analytic extension for $f_k^{(1)}$: A mode that covers more than just region I now. But as defined, $f_k^{(1)}$ doesn't cover of spacetime, since the modes are zero in region IV. Then, we need to include the $f_k^{(2)}$. What we have to do is come up with a version of the $f_k^{(2)}$ that has the {\it same form} as the r.h.s. of (\ref{funI}) in terms of the Minkowski coordinates evaluated on IV. If we simply evaluate it in region IV (again with $\omega_k=k$) 
\bee
\sqrt{4\pi\omega_k}f_k^{(2)} = e^{i\omega_k(\gamma+\delta)}=(e^{a(\gamma+\delta)})^{i\omega_k/a}=a^{i\omega_k/a}(-t-x)^{i\omega_k/a}.
\eee
This doesn't match with the r.h.s. of (\ref{funI}); we need something of the form $a^{i\omega_k/a}(-t+x)^{i\omega_k/a}$. What does work is
 \bee
\sqrt{4\pi\omega_k}f_{-k}^{(2)^*} =e^{-i\omega_k(\gamma-\delta)}=(e^{-a(\gamma-\delta)})^{i\omega_k/a}= a^{i\omega_k/a}(-1)^{i\omega_k/a}(-t+x)^{i\omega_k/a}.
\eee
So if we add the two modes in the following combination 
\bee
f_k^{(1)}+(-1)^{-i\omega_k/a}{f_{-k}^{(2)}}^*= a^{i\omega_k/a}(-t+x)^{i\omega_k/a}, 
\eee
we have a function that covers both region I and IV, that is, an analytic extension for $f_{k}^{(1)}$ to the other side using that the analyticity of $(-t+x)^{i\omega_k/a}$ in the complex plane. 

Note that in terms of the variable $v=t-x$ the function $(-t+x)^{i\omega_k/a}=(-v)^{i\omega_k/a}=\exp\left(i\frac{\omega_k}{a}\ln (-v)\right)$ is singular at the origin, that is the reason why the analytic extension is being used: Thinking of $v=t-x$ as a complex variable one can choose if the function is going to be analytic by going {\it over} or {\it under} the origin in a complex plane. Both results are different, and this is expressed in the ambiguity to define $(-1)^{i\omega_k/a}=e^{\pm\pi\omega_k/a}$ which we need to specify now. 

But first let us recap: We look for a complete basis to expand scalar fields in Minkowski spacetime, but we want a basis built using the Rindler modes. We found a way to extend a Rindler mode from region I to region IV but there is an ambiguity in the extension. We now develop a criteria to solve the ambiguity. 

Let's establish the following statement

\begin{center}
{\it A function is a combination of only positive frequency modes if and only if \\ as a complex function it is analytic and bounded in the lower complex plane.}
\end{center}

First, consider the usual Minkowski observer's positive frequency modes $f_k=A_ke^{-i\omega_k(t-x)}$ with $\omega_k>0$. Thought as a function in the complex plane for the variable $v=t-x$, the function $e^{-i\omega_k v}$ is analytic and bounded in the lower half complex plane: Note that in the lower half plane $\text{Im} (v) \leq 0$ and the function, $e^{\omega_k \text{Im}(v)}e^{-i\omega_k \text{Re}(v)}$,  decreases exponentially as we go to infinity in any direction of the lower half complex plane. Now, if we consider a function that is a combination of positive frequency modes (each mode weighted by a function $\tilde h(\omega_k)$)
\bee
h(v)=\int_0^\infty d\omega_k e^{-i\omega_k v}\tilde h(\omega_k),
\eee
it also has the same property, it is analytic and bounded in the lower half of the complex plane. But note that this property is in fact a sufficient condition: If in the combination there is a single negative frequency mode, then the function blows up in the lower half of the complex plane. Therefore, being positive frequency for a real variable is equivalent to being analytic and bounded in the lower complex plane. Thus the statement holds.

\vspace{0.1cm}

Now we have an elegant criterion to choose the analytic extension of our modes. We extend $f^{(1)}_k$ to the region IV and make a choice on the extension such that 
\bee
f_k^{(1)}+(-1)^{-i\omega_k/a}{f_{-k}^{(2)}}^*=a^{i\omega_k/a}\exp\left(i\frac{\omega_k}{a}\ln (-v)\right),
\eee
 becomes a combination of only positive frequency modes, or, equivalently such that as a complex function of $v=t-x$  it is analytic and bounded in the lower half complex plane. 

\begin{figure}[h]
\centering
\includegraphics[scale=0.5]{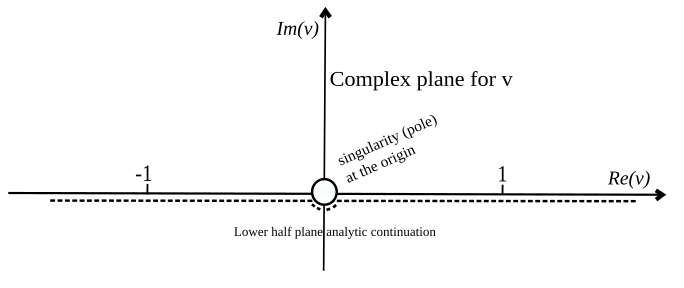}
\caption{The dashed line represents the path over which the function $\ln{(-v)}$ is extended from the right to the left side of the real line. In particular $-1=e^{i(-\pi)}=e^{-i\pi}$.}
\label{figpole2}
\end{figure}

The complex logarithm is singular at the origin and is also a multivalued function, this is clear in polar coordinates $\ln (-v)=\ln (r e^{i\theta})+\ln(e^{i(2n-1)\pi})=\ln r +i\theta+i(2n-1)\pi$, with $n\in \mathbb{Z}$. Then, by excluding the origin, taking $n=0$, and restricting $\theta$ to the interval $\theta\in[0,-\pi]$, we define it as an analytic and single-valued function covering the {\it lower half complex plane}; with this choice we are essentially ``continuing'' the function along the lower half plane.\footnote{In the language of complex variable this is equivalent to choosing a {\it branch cut} in the upper half complex plane that starts at origin.} In agreement with this we must select the minus one as $(-1)^{-i\omega_k/a}=(e^{-i\pi})^{-i\omega_k/a}=e^{-\pi\omega_k/a}$, see figure \ref{figpole2}.

Therefore, we define a new set of functions which is a combination of positive frequency modes as
\be
h_k^{(1)} = \frac{1}{\sqrt{2\sinh(\frac{\pi\omega_k}{a})}} \left(e^{\pi\omega_k/2a}f_k^{(1)}+e^{-\pi\omega_k/2a}{f_{-k}^{(2)}}^*\right), \label{ext1} 
\ee
where we included an overall factor for normalization (as you're encouraged to verify in the exercises). This, finally, is the analytic extension of the $f_k^{(1)}$ modes. We carried out this procedure assuming $k>0$, but you may verify that with $k<0$ one arrives at the same linear combination. 

Now, the set  $f_k^{(1)}$ and  $f_k^{(2)}$, together with their conjugates, form a complete set. We have just built $h_k^{(1)}$ by extending the $f_k^{(1)}$ modes, to have a complete basis we need to analytically extend the $f_k^{(2)}$ modes too.  


We start, like before, by expressing $f_k^{(2)}$ in terms of Minkowski coordinates in region IV. Assuming $k>0$, 
\bee
\sqrt{4\pi\omega_k}f_{k}^{(2)} = e^{i\omega_k(\gamma+\delta)} = (e^{a(\gamma+\delta)})^{i\omega_k/a}=a^{i\omega_k/a}(-t-x)^{i\omega_k/a},
\eee 
since in region IV, $e^{a(\gamma+\delta)}=a(-t-x)$. The $f_k^{(2)}$ vanish in region I, then, the extension needs a version of the $f_k^{(1)}$ (that looks the same as what we have above) that we can add to this. For region I, $e^{a(\eta+\xi)}=a(t+x)$, which is close to the above expression, save for a factor of $-1$. We try 
\begin{align*}
\sqrt{4\pi\omega_k}f_{-k}^{(1)^*} &= e^{i\omega_k(\eta+\xi)} \\
&= a^{i\omega_k/a} (t+x)^{i\omega_k/a} \\
&= a^{i\omega_k/a} (e^{-i\pi}(-t-x))^{i\omega_k/a} \\
& = a^{i\omega_k/a}e^{\pi\omega_k/a}(-t-x)^{i\omega_k/a}. 
\end{align*}
So in a similar way to (\ref{ext1})  we define a normalized extension of the $f_k^{(2)}$
\be
h_k^{(2)} = \frac{1}{\sqrt{2\sinh(\frac{\pi\omega_k}{a})}} \left(e^{\pi\omega_k/2a}f_k^{(2)} +e^{-\pi\omega_k/2a}{f_{-k}^{(1)}}^*\right).
\ee
Thus, the field expansion in the whole Minkowski spacetime can be carried on using either the standard modes $(\varphi_{\vec{k}},\varphi^*_{\vec{k}})$ or the new set of modes $(h_k^{(1)},h_k^{(2)},h_k^{(1)^*},h_k^{(2)^*})$ just developed. The scalar field in the new modes is
\begin{align*}
\phi = \int dk \left(\h{c}_k^{(1)} h_k^{(1)}+\h{c}^{(1)\D}_k {h_k^{(1)}}^*+ \h{c}_k^{(2)} h_k^{(2)}+\h{c}^{(2)\D}_k {h_k^{(2)}}^*\right),
\end{align*}
the explicit expression in term of the Rindler modes
\begin{align*}
\phi &= \int dk \frac{1}{\sqrt{2\sinh(\frac{\pi\omega_k}{a})}} \left\{\h{c}_k^{(1)} \left(e^{\pi\omega_k/2a}f_k^{(1)}+e^{-\pi\omega_k/2a}{f_{-k}^{(2)}}^*\right) + \h{c}^{(1)\D}_k \left( e^{\pi\omega_k/2a}{f_k^{(1)}}^*+e^{-\pi\omega_k/2a}{f_{-k}^{(2)}}\right)  \right. \\
&\quad\quad\quad\quad\quad\quad\quad\quad\quad+\left.\h{c}_k^{(2)}  \left(e^{\pi\omega_k/2a}f_k^{(2)} +e^{-\pi\omega_k/2a}{f_{-k}^{(1)}}^*\right) + \h{c}^{(2)\D}_k  \left(e^{\pi\omega_k/2a}{f_k^{(2)}}^* +e^{-\pi\omega_k/2a}{f_{-k}^{(1)}}\right)  \right\}.
\end{align*}
In this expansion, the $\h{c}_k$'s are associated to positive frequency Minkowski modes and play the r\^ole of annihilation operators, while the $\h{c}_k^{\D}$'s are associated to negative frequency Minkowski modes and then play the r\^ole of creation operators in this basis. Therefore, the Minkowski vacuum $\ket{0_M}$ for an observer using the $h_k$ modes will satisfy 
\be
\h{c}_k^{(1)}\ket{0_M}=\h{c}_k^{(2)}\ket{0_M}=0,\quad\quad \text{for all $k$.}
\ee
We now recall that the natural expansion a Rindler observer uses for $\phi$ is  
\bee
\phi = \int dk\left(\h{b}_kf_k^{(1)} + \h{b}_k^{\D} {f_k^{(1)}}^* \right).
\eee
This, of course, is only defined in region I. We can equate the $\phi$ expansion in $h_k$ to this expression, remembering that the equality is only valid in region I. Then, because in region I the set $(f^{(1)}_k, f_k^{(1)^*})$ form a basis, the equality implies the matching of the coefficients of the $f_k^{(1)}$ functions, and we have 
\ba
\h{b}_k = \frac{1}{\sqrt{2\sinh(\frac{\pi\omega_k}{a})}} \left(e^{\pi\omega_k/2a}\h{c}^{(1)}_k + e^{-\pi\omega_{k}/2a}\h{c}_{-k}^{(2)\D}\right),
\ea
where we used that $\omega_{-k}=\omega_k$. This is the Bogoliubov transformation of operators defined for different bases covering the Rindler wedge. Note that thanks to the analytic extension procedure we reached it without the explicit computation of complex integrals.

Now, with the Bogoliubov transformation at hand extracting some physics is easy. In general we may compute how any quantum state in the rest frame is described in the accelerated one and vice versa. But we can go straight to a surprising physical effect if instead we calculate the number of particles that an accelerating observer will see in the Minkowski vacuum $\ket{0_M}$  
\begin{align}
\braket{\h{n}_{{b_k}}} &= \braket{0_M|\h{b}^{\D}_k\h{b}_k|0_M} \nonumber \\
& = \frac{1}{2\sinh(\frac{\pi\omega_k}{a})} \Braket{0_M|\left(e^{\pi\omega_k/2a}\h{c}^{(1)\D}_k + e^{-\pi\omega_{k}/2a}\h{c}_{-k}^{(2)}\right)\left(e^{\pi\omega_k/2a}\h{c}^{(1)}_k + e^{-\pi\omega_{k}/2a}\h{c}_{-k}^{(2)\D}\right)|0_M} \nonumber \\ 
& = \frac{1}{2\sinh(\frac{\pi\omega_k}{a})}  \Braket{0_M|e^{-\pi\omega_k/a}c_{-k}^{(2)}c_{-k}^{(2)\D}|0_M}\nonumber \\
& = \frac{ e^{-\pi\omega_k/a}}{e^{\pi\omega_k/a}-e^{-\pi\omega_k/a}} \delta(0)\nonumber\\
& = \frac{1}{e^{2\pi\omega_k/a}-1}\delta(0).
\end{align}
The $\delta(0)$ comes from the fact that our basis modes are not square-integrable; if we restrict the size of spacetime to be finite, the factor would instead be a constant. 

Then, we have found
\be
\braket{\h{n}_{{b_k}}} \sim  \frac{1}{e^{2\pi\omega_k/a}-1}.
\ee
 
What we learn from the above equation is that in flat spacetime, when an inertial observer describes a vacuum state, an accelerating observer describes particles.  Furthermore, for constant acceleration, those particles have a very specific distribution that corresponds to a Planckian thermal radiation. The characteristic temperature is the so called Unruh temperature
\be
T = \frac{a}{2\pi}, 
\ee
which is linear in the proper acceleration of the observer. Therefore, just because of the acceleration, an accelerated object is in a thermal bath with an Unruh temperature.

Putting the constants back in, we get a sense of the weakness of the Unruh effect
\be
T = a\cdot\frac{\hbar}{2\pi ck_B}\approx a\cdot 4 \times 10^{-21}\left[K\cdot s^2/m\right]. 
\ee
Thus, to heat up your cup of tea you need to give it a proper acceleration of about $9\times 10^{22}\left[m/s^2\right]$. 

The huge scale of the acceleration almost fades out the hope of finding a direct experimental signal of the Unruh effect. However, analogue experimental setups and theoretically related questions makes of this striking physical effect a fueling topic in current research. We develop on this in the following.

\section{Discussion}
\label{discussion}

The Unruh effect was described in 1976, just a couple of years after Hawking radiation for black holes. The mathematics behind both effects is the same but what changes is the spacetime background and then the physical interpretation of each phenomenon. In this sense the Unruh effect is simpler than the Hawking effect because it relies on flat spacetime. However,  Hawking radiation has proven to be far more interesting as it changed the classical understanding of black holes at the time, and also because it prompted several questions regarding the foundations of physics.\footnote{A few questions are reverberating till today: What is the fate of information falling into black holes? What is the statistical description of black hole entropy? What happens at the end of black hole evaporation? (For a recent review see \cite{Unruh:2017uaw}). In the words of Kip Thorne at the Hawking's burying ceremony ``We remember Isaac Newton for answers, we remember Hawking for questions.'' }

Besides the mathematical aspect, the connection between both effects can be pushed further. There is a bold physical interplay between the Hawking and Unruh effect. We briefly sketch it.  Consider a static observer close to a black hole horizon. To keep the position this observer experiences a high proper acceleration; the closer the observer is to the horizon the higher the acceleration. In fact the spacetime in the neighborhood of the observer is very well approximated by a Rindler metric. Thus, for a static observer close enough to the horizon the radiation he measures is basically Unruh radiation due to its high proper acceleration rather than a black hole radiation. In fact, by reversing the logic, an indirect way to recover the Hawking radiation is to consider the Unruh radiation for near horizon static observers and use the redshift of the radiation to an observer far from the horizon. In this way the Hawking temperature is correctly recovered. 


\vspace{0.3cm}

As you may have noticed in these notes, finding the Unruh temperature was the icing on the cake but at the same time it is just the tip of the iceberg. One deep thing we learned is that ``particles'' (or field excitations) are not frame independent. Given a particular quantum state for the fields, it is possible for non inertial observers to detect field excitations where inertial observers do not.

One can make the parallel of ``particles'' in quantum field theory with fictitious forces in Newtonian mechanics. Both phenomena emerge for non inertial frames. However, there are interesting historical differences. In the case of fictitious forces, the phenomenon becomes trivial once it is realized that Newtonian mechanics is replaced by general relativity. The covariant aspect of general relativity, the fact that the equations governing the mechanics are written the same fashion regardless the frame, trivialize the appearance of fictitious forces.\footnote{Note that Newtonian mechanics may also be written in a covariant language, e.g. Newton-Cartan theory, where Galilean symmetry plays the role of Lorentz symmetry. So we emphasize that it is the use of a covariant language, and not the full general relativity, what clears the issue.} While in the case of field excitations (``particles'') in quantum field theory, as we observe in the derivation of the Unruh effect, we did not use a covariant language and thus the appearance or not of field excitations is an issue far from being trivial. The Fock space and annihilation/creation operators used to build it rely on particular frames.

In fact, it would be desirable to implement covariance for quantum field theory. That is, to use a more abstract language such that the equations are written in the same fashion independently of the frame. This can be achieved in some formulations of quantum field theory in curved spacetimes (see section \ref{pcr}). In those, it becomes clear that the appearance of particles, as in the Unruh effect, is a feature of specific setups of quantum states and frames. 

However, this is not the end of the story. Besides covariance, which just stands for using tensorial language to write the expression in the theory, general relativity introduces the gravitational field and the equivalence principle. Those new ingredients are by no means well-understood in the context of quantum field theory and claim for a quantum gravity theory: The theoretical conundrum of our times. 

\vspace{0.3cm}

\subsection{A Peek into Current Research}
\label{pcr}

To close, we make some comments on different avenues of the Unruh effect. Being an old enough subject there are several paths that have been followed so we highlight just three of them that have been active in recent years. The cited literature in the following is meant as a particular opening in each direction.

Before that it is worth to point out that the interpretation of the Unruh effect as a thermal bath state for accelerated observers is controversial. The thermal response is straightforward for an accelerated point-like detector or for a collection of them, however this is a rather simplified model. As far as the picture is refined by considering real accelerating labs of finite size, the description of the vacuum state in the accelerated frame encounter several subtleties. For a critical account of the thermal bath interpretation we refer to \cite{Buchholz:2015fqa}.
 
The Unruh effect in the context of {\bf analogous gravity} \cite{Barcelo:2005fc} deserves special attention. The main observation is that there are physical systems in condensed matter with an intrinsic maximum speed for the excitations of the system to propagate. In some of those systems the excitations propagate on a background that can adopt non trivial configurations with ``horizons'' or directional membranes for the propagation. An effective quantum field theory in curved spacetime may be applied and spontaneous radiation from the horizons is predicted.  This is a very active field in the recent years with a high observational prospect for an analog of Hawking radiation \cite{Weinfurtner:2013zfa}.

As stated in the introduction the Unruh effect is a natural motivation to explore {\bf quantum field theory in curved spacetimes}. The reason is that the technical machinery used may be generalized to curved spacetimes. The main assumption here, and the reason why this is not a theory of {\it quantum gravity}, is to consider an arbitrary curved spacetime, satisfying Einstein's equations, without quantizing the field $g_{\mu\nu}$. Then, known tools to describe quantum fields in Minkowski spacetime are generalized to background spacetimes that lack any a priori symmetry. Note that in this approach the gravity field can still be treated as a quantum perturbation over the background. However, this approach is not complete because it does not deal with the backreaction of the quantum fields on spacetimes. A clear example of the importance of considering the backreaction is the description of an ``eternally radiating'' black hole. Backreaction considerations would tells us how the black hole actually shrinks. 

Having said that, still the formulation of QFT in curved spacetime is a very challenging and powerful field. General relativity, as a classical theory, is an excellent approximation for all the observed spacetime. Therefore, to understand how quantum fields are formulated on it may help unveil new physical phenomena, to find the limitations of our physical descriptions, or even to draw a path for a full quantum gravity theory. 

As an old enough subject there are many, even classical, references. Here we point out the current mathematical approach named algebraic quantum field theory in curved spacetime \cite{Fredenhagen:2015utr} and \cite{Fewster:2015kua} (references there are extensive on classical works and current research).

Finally, another promising avenue to explore in the context of the Unruh effect is the search of {\bf quantum gravity} modifications to the thermal spectrum as measured by an accelerating observer. Naively, for high accelerations, and corresponding high Unruh temperatures, the expectation is that other physical effects play a r\^ole. If the classical description of spacetime is going to be refined by a quantum gravity description, the Unruh effect will require corrections, i.e., quantum gravity corrections even for approximate flat spacetimes. An interesting analogy is to think of a thermometer in a fluid medium moving at a constant speed. For high speeds the interaction of the fluid particles and the thermometer will affect the temperature measurements. Thus the microscopic structure plays a r\^ole. The same idea applies for an accelerated observer measuring the thermal radiation in a spacetime described as a coarse-grained structure. In this context ideas of a fundamental maximal acceleration (e.g. Planck acceleration) or spontaneous black hole creation (Hawking-Page phase transition) may be considered. This avenue is certainly the most promising one because it is not fully explored and also because it is a way to bring back quantum gravity proposals to the reality of physical predictions. 

In this direction recent proposals have used the so-called Unruh-DeWitt detector approach. In particular is has been explored the modification of the spectrum due to quantum gravity models presenting dimensional flow at short scales \cite{Alkofer:2016utc}, or general treatments considering breaking of the Lorentz symmetry due to short length parameters \cite{Carballo-Rubio:2018zll}.

\newpage

\section*{Acknowledgments}

The authors would like to thank Rodrigo Eyheralde for reading and commenting on an early version of these notes. NV is grateful to Mart\'in Gilabert for a helpful comment in section 2.3. EF is partially funded by FONDECYT grant \#11150467. The Centro de Estudios Cient\'ificos (CECs) is funded by the Chilean Government through the Centers of Excellence Base Financing Program of Conicyt.

\section{Exercises}  

{\scriptsize

\begin{enumerate}
\item \label{timederivative} Show that the inner product we defined (on the solution space of the harmonic oscillator equation) is conserved in time. 
\vspace{-0.3cm}
\item Consider $f(t)$ and $g (t)$ two solutions of $\ddot{h}+\omega^{2}(t)h=0$. Because it is a second order equation they are related by 
\begin{equation}\label{eq2}
g=\alpha f+\beta f^*,\quad\quad\quad\quad\quad \quad \alpha,\beta\in \mathbb{C}.
\end{equation}
Show that normalization implies 
\begin{equation}
\vert \alpha \vert^{2} -\vert \beta \vert^{2} =1.
\end{equation}

\vspace{-0.3cm}
\item From (\ref{eq2}), derive the Bogoliubov transformation among the operators $\hat{a},\ \hat{a}^{\dagger},\ \hat{b}$, and $\hat{b}^{\dagger}$ defined through
\ba
\h{x}&=&\h{a}f(t)+\h{a}^{\dagger}f^*(t)\\
&=&\h{b}\, g(t)+\h{b}^{\dagger}g^*(t).
\ea
Show that if $\vert 0\rangle _{f}$ is the state annihilated by $\hat{a}$, and $\hat{N}_{g}$ is the number operator given by $\hat{N}_{g}=\hat{b}^{\dagger}\hat{b}$, then
\begin{equation}\nonumber
\langle 0\vert _{f} \hat{N}_{g}\vert 0 \rangle _{g}=\vert \beta \vert^{2}.
\end{equation}
Comment on this result.
\vspace{-0.3cm}
\item Show that positive and negative frequency modes are orthogonal, i.e. $\braket{\varphi_{\vec{k}}|\varphi_{\vec{q}}^*}=0$. Also show that $\braket{\varphi^*_{\vec{k}}|\varphi_{\vec{q}}^*}=-\delta^{(n-1)}(\vec{k}-\vec{q})$. 
\vspace{-0.3cm}
\item Check that the momentum and energy of the state
\begin{equation}
\hat{a}_{\vec{k}}^{\dagger}\vert 0\rangle,
\end{equation}
are $\hslash \vec{k}$ and $\hslash \omega$, respectively.
\vspace{-0.3cm}
\item Show that if we plug  the explicit versions of $\varphi_{\vec{k}}$ into the fields $\phi$ and $\pi$, we get the commutation relation $[\phi(t,\vec{x}),\pi(t,\vec{y})]=i\hbar\delta^{(n-1)}(\vec{x-y})$ (assuming we know how the $\h{a}_{\vec{k}}$ commute). 
\vspace{-0.3cm}
\item Do the necessary calculations to arrive at (\ref{qftH}). 
\vspace{-0.3cm}
\item What would be the form of the Euler-Lagrange equations for fields in curved spacetime? Deduce (\ref{motion3}), and the conjugate momentum to $\phi$. \textit{Hint:} consider the identity $\nabla_{\mu}V^{\mu}=\frac{1}{\sqrt{-g}}\pd_{\mu}(\sqrt{-g}V^{\mu})$, and remember that the covariant derivative is metric-compatible. 
\vspace{-0.3cm}
\item Sketch the paths of observers moving in Minkowski spacetime with constant accelerations $a$, $2a$, $\infty$, $0$. 
\vspace{-0.3cm}
\item Invert the definition of Rindler coordinates to get explicit expressions for $\xi$ and $\eta$.
\vspace{-0.3cm}
\item Verify that in Rindler coordinates, $\xi=\frac{1}{a}\ln(\frac{a}{\alpha})$ (constant) for an observer moving with constant acceleration $\alpha$. Additionally, check that the proper time for a Rindler observer is proportional to the coordinate $\eta$. 
\vspace{-0.3cm}
\item Verify that $\pd_{\eta}$ is a Killing vector field. 
\vspace{-0.3cm}
\item Show that $\pd_{\gamma}$ is timelike in region IV. 
\vspace{-0.3cm}
\item Verify that the $h_k$ modes are properly normalized. 

 \end{enumerate}

}

\end{document}